\documentclass[conference]{IEEEtran}

\usepackage{listings}
\lstdefinestyle{shell}{
  language=sh,
  basicstyle=\ttfamily\small,
  numbers=left,
  numberstyle=\tiny\color{gray},
  stepnumber=1,
  numbersep=10pt,
  backgroundcolor=\color{white},
  showspaces=false,
  showstringspaces=false,
  showtabs=false,
  frame=single,
  rulecolor=\color{black},
  tabsize=2,
  captionpos=b,
  breaklines=true,
  breakatwhitespace=false,
  escapeinside={\%*}{*)},
  morekeywords={ACTION_DOWN, ACTION_UP, ACTION_MOVE}
}
\usepackage[]{algorithm2e}
\SetKwInput{KwData}{Input}
\SetKwInput{KwResult}{Output}
\usepackage{tcolorbox}
\usepackage{multirow}
\usepackage{multicol}
\usepackage{amsmath}
\usepackage{pifont}
\usepackage{colortbl}
\usepackage{balance}
\usepackage{tabularx}
\usepackage{url}
\usepackage{ragged2e}
\usepackage{booktabs,caption}
\usepackage[flushleft]{threeparttable}

\usepackage{enumitem}
\usepackage{listings}
\usepackage{graphicx}
\usepackage{caption}
\usepackage{subcaption}
\usepackage{xcolor}
\definecolor{dkgreen}{rgb}{0.0, 0.5, 0.0}
\usepackage{mathptmx}

\newcommand{\perfailure}{38\%}
\newcommand{\percrash}{44\%}
\newcommand{\perscenario}{17\%}
\newcommand{\crblue}[1]{\textcolor{black}{#1}}

\newcommand{\OurComment}[1]{}
\newcommand{\Space}[1]{}
\newcommand{\Num}[1]{#1} 

\newcommand{\Def}[2]{\expandafter\newcommand\csname rmk-#1\endcsname{#2}}
\newcommand{\Use}[1]{\csname rmk-#1\endcsname}

{\makeatletter
 \gdef\jonmark{%
   \expandafter\ifx\csname @mpargs\endcsname\relax 
     \expandafter\ifx\csname @captype\endcsname\relax 
       \marginpar{\color{blue}{jon~}}
     \else
       \color{blue}{jon~}
     \fi
   \else
     \color{blue}{jon~}
   \fi}
 \gdef\jon{\@ifnextchar[\jon@lab\jon@nolab}
 \long\gdef\jon@lab[#1]#2{{\bf [\jonmark \color{blue}{#2} ---{\sc #1}]}}
 \long\gdef\jon@nolab#1{{\bf [\jonmark \color{blue}{#1}]}}
}

{\makeatletter
 \gdef\aaronmark{%
   \expandafter\ifx\csname @mpargs\endcsname\relax 
     \expandafter\ifx\csname @captype\endcsname\relax 
       \marginpar{\color{purple}{aaron~}}
     \else
       \color{purple}{aaron~}
     \fi
   \else
     \color{purple}{aaron~}
   \fi}
 \gdef\aaron{\@ifnextchar[\aaron@lab\aaron@nolab}
 \long\gdef\aaron@lab[#1]#2{{\bf [\aaronmark \color{purple}{#2} ---{\sc #1}]}}
 \long\gdef\aaron@nolab#1{{\bf [\aaronmark \color{purple}{#1}]}}
}

\newcommand{\cmark}{\ding{51}}%
\newcommand{\xmark}{\ding{55}}%

\newcommand{\etal}{et al.}

\usepackage{url}
 
\newcommand{\numMonkeyDetected}{38\xspace}
\newcommand{\numApeDetected}{61\xspace}

\newcommand{\numHumanoidDetected}{26\xspace}

\newcommand{\numMonkeyRecorded}{10\xspace}
\newcommand{\numApeRecorded}{5\xspace}
\newcommand{\numHumanoidRecorded}{3\xspace}

\newcommand{\ape}{APE\xspace}
\newcommand{\recordreplay}{R\&R\xspace}
\newcommand{\recordreplayFull}{record-and-replay\xspace}

\newcommand{\aig}{AIG\xspace}
\newcommand{\reran}{RERAN\xspace}
\newcommand{\replaykit}{ReplayKit\xspace}
\newcommand{\sara}{SARA\xspace}

\Def{sleepyline.cause}{Sleep Reasons}

\Def{config.Count}{Configs. (\%)}
\Def{configs.proportion}{configs.proportion}

\Def{failureID.Uniq.Count}{Failures (\%)}
\Def{failureID.Uniq.Proportion}{failures.proportion}

\Def{test.count}{Tests (\%)}
\Def{test.proportion}{tests.proportion}

\Def{FLAGGED_API}{API}
\Def{RUNNABLE_CALLABLE_START}{ThreadBegin}
\Def{MONITORENTER}{EnterSyncBlock}
\Def{MONITOREXIT}{ExitSyncBlock}
\Def{SYNCHRONIZED_METHOD_ENTER}{EnterSyncMethod}
\Def{SYNCHRONIZED_METHOD_EXIT}{ExitSyncMethod}
\Def{java/lang/System.currentTimeMillis()J}{System.currentTimeMillis}
\Def{java/lang/Thread.currentThread()Ljava/lang/Thread;}{Thread.currentThread}
\Def{java/lang/Object.notifyAll()V}{Object.notifyAll}
\Def{java/io/OutputStream.flush()V}{OutputStream.flush}
\Def{java/nio/ByteBuffer.allocate(I)Ljava/nio/ByteBuffer;}{ByteBuffer.allocate}
\Def{java/nio/ByteBuffer.array()[B}{ByteBuffer.array}
\Def{java/nio/channels/Selector.wakeup()Ljava/nio/channels/Selector;}{Selector.wakeup}

\newboolean{showcomments}

\setboolean{showcomments}{true}

\ifthenelse{\boolean{showcomments}}
  {\newcommand{\nb}[2]{
    \fbox{\bfseries\sffamily\scriptsize#1}
    {\sf\small$\blacktriangleright$\textit{#2}$\blacktriangleleft$}
   }
   
  }
  {\newcommand{\nb}[2]{}
   
  }

\makeatletter 
\newcommand{\linebreakand}{%
  \end{@IEEEauthorhalign}
  \hfill\mbox{}\par
  \mbox{}\hfill\begin{@IEEEauthorhalign}
}
\makeatother

\begin{document}

\title{Can You Mimic Me? Exploring the Use of Android Record \& Replay Tools in Debugging}

\author{
    \centering
    \IEEEauthorblockN{Zihe Song}
    \IEEEauthorblockA{
    \textit{University of Texas at Dallas}\\
    Richardson, TX, USA \\
    zihe.song@utdallas.edu}
    \and
    \IEEEauthorblockN{S M Hasan Mansur}
    \IEEEauthorblockA{
    \textit{George Mason University}\\
    Fairfax, VA, USA \\
    smansur4@gmu.edu}
    \and
    \IEEEauthorblockN{Ravishka Rathnasuriya}
    \IEEEauthorblockA{
    \textit{University of Texas at Dallas}\\
    Richardson, TX, USA \\
    Ravishka.Rathnasuriya@utdallas.edu}
    \and
    \IEEEauthorblockN{Yumna Fatima}
    \IEEEauthorblockA{
    \textit{George Mason University}\\
    Fairfax, VA, USA \\
    yfatima@gmu.edu}
    
    \linebreakand
    
    \IEEEauthorblockN{Wei Yang}
    \IEEEauthorblockA{
    \textit{University of Texas at Dallas}\\
    Richardson, TX, USA \\
    wei.yang@utdallas.edu}
    \and
    \IEEEauthorblockN{Kevin Moran}
    \IEEEauthorblockA{
    \textit{University of Central Florida}\\
    Orlando, FL, USA \\
    kpmoran@ucf.edu}
    \and
    \IEEEauthorblockN{Wing Lam}
    \IEEEauthorblockA{
    \textit{George Mason University}\\
    Fairfax, VA, USA \\
    winglam@gmu.edu}
}

\maketitle

\begin{abstract}
Android User Interface (UI) testing has emerged as an important and prevalent research topic due to the ubiquity of apps and the unique challenges faced by developers in this software domain. One popular topic of research that aims to facilitate both manual and automated UI testing and debugging processes is record and replay (R\&R) tools. These tools allow for the recording of UI actions to facilitate the execution of test scenarios and the replay of various types of bugs. R\&R tools typically support three main settings: (i) UI regression testing via R\&R of feature-based execution scenarios, (ii) R\&R of non-crashing functional bugs (e.g., in crowdsourced settings), and (iii) R\&R of crashing bugs. 
Despite the progress made in research related to R\&R tools, prior work examined only the effectiveness of these tools in disparate or fragmented settings. As such, the research community currently lacks a \textit{comprehensive} examination of the effectiveness of existing tools across their common use cases and the potential key limitations that emerge.

We address this current gap in knowledge by conducting a thorough empirical study on using R\&R tools to manually record and replay feature-based user scenarios, non-crashing failures, and crashing bugs. Additionally, we explore the possibility of using R\&R tools in conjunction with automated input generation~(AIG) tools to automatically record and replay crashing bugs. Our study context includes one industrial and three academic R\&R tools, 34 user scenarios from 17 apps, 90 non-crashing failures from 42 Android apps, and 31 crashing bugs from 17 Android apps. Our results illustrate that \perscenario{} of user scenarios, \perfailure{} of non-crashing failures, and \percrash{} of crashing bugs are not able to be reliably recorded and replayed, with the most prevalent reasons for non-replayability being action interval resolution, incompatibility related to APIs, and limitations in Android tooling. Our findings reveal important\Space{ future} research directions for R\&R tools to facilitate their practical application and adoption.
\end{abstract}


\section{Introduction}
\label{sec:intro}
In recent years, \recordreplayFull (\recordreplay) tools have become one of the most popular types of automated testing tools for mobile and web applications, facilitating the precise capture of UI actions in detailed traces. With continued development by both researchers and developers, \recordreplay tools have evolved from the record and replay of low-level event sequences~\cite{reran} to capturing system-level events~\cite{VALERA}, and most recently, operating upon screen recordings of mobile apps to replay depicted actions and gestures~\cite{V2S}. \crblue{However, beyond traditional test automation, these tools also hold substantial potential for debugging, particularly in reproducing bugs that are difficult to diagnose using conventional debugging techniques.}

\crblue{
Bug reproduction plays a crucial role in software debugging, as it allows developers to verify issues, understand the conditions under which they occur, and analyze how different components interact to cause failures. While stack traces and logs provide snapshots of system states at the time of failure, they often fail to capture the full sequence of actions leading to the bug. This limitation is especially problematic for issues arising from race conditions, timing-dependent behaviors, or non-deterministic executions, where a precise reproduction of the user’s actions is necessary for effective debugging. \recordreplay tools, by enabling faithful re-execution of UI interactions, have the potential to fill this gap and support developers in both debugging and regression testing, ensuring that bug fixes resolve the issue without introducing new problems.}

\begin{table*}[t]
    \centering
    \caption{This table highlights the key differences between our study and prior relevant studies. 
    Our study evaluates the reproducibility of \recordreplay tools in four different use cases: (1) common user scenarios, (2) failures from bug reports, (3) crashes from bug reports, and (4) crashes from \aig tools. 
    We also evaluate the reproducibility of crashes from \aig tools using just the \aig tools themselves. 
    \cmark{} and \xmark{} denote that the study does and does not, respectively, evaluate a particular use case.}
    \begin{tabular}{l|l|l|c|c|c|c|c}
        & & \multicolumn{1}{c|}{\textbf{Tool}} & \multicolumn{1}{c|}{\textbf{User}} & \multicolumn{1}{c|}{\textbf{Bug Report}} & \multicolumn{3}{c}{\textbf{Crashes From}} \\
        \multicolumn{1}{c|}{\textbf{Study}} & \multicolumn{1}{c|}{\textbf{Venue}} & \multicolumn{1}{c|}{\textbf{Basis}} & \multicolumn{1}{c|}{\textbf{Scenarios}} & \multicolumn{1}{c|}{\textbf{Failures}} & \multicolumn{1}{c|}{\textbf{Bug Reports}} & \multicolumn{1}{c|}{\textbf{\aig Tools}} & \multicolumn{1}{c}{\textbf{\aig w/o \recordreplay Tools}}\\ \hline
        Lam~\etal~\cite{lam17:record} & ESEC/FSE'17 Industry & \recordreplay & \cmark & \xmark & \xmark & \xmark & \xmark \\
        Su~\etal~\cite{Su2021WS} & ESEC/FSE'21 & \aig & \xmark & \xmark & \xmark & \cmark & \xmark \\
        Liu~\etal~\cite{liu2023understanding} & SETTA'23 & \aig & \xmark & \xmark & \xmark & \cmark & \xmark \\
        Xiong~\etal~\cite{xiong2023empirical} & ISSTA'23 & \aig \& \recordreplay & \xmark & \cmark & \xmark & \xmark & \xmark \\
        Our study & - & \aig \& \recordreplay & \cmark & \cmark & \cmark & \cmark & \cmark \\
    \end{tabular}
    \vspace{-1em}
    \label{tab:study:compare}
\end{table*}

\crblue{
Despite this potential, prior research has primarily focused on evaluating \recordreplay tools in terms of their effectiveness in replaying general UI interactions rather than their ability to reproduce bugs. Existing studies typically assess these tools using controlled scenarios that reflect common user behaviors rather than complex failure cases.}
For example, \sara~\cite{sara} was evaluated only on common user scenarios across popular applications. Our prior study~\cite{lam17:record} of \recordreplay tools also focused solely on common user scenarios. This focus has likely contributed to recent reported findings~\cite{Linares-Vasquez:ICSME'17} that automated tools are largely passed over in favor of manual testing techniques by professional and open-source Android app developers. \crblue{
Compared to general UI interaction replay, bug reproduction presents unique challenges: it often involves unusual action sequences, interactions with sensors, and conditions that may destabilize the application itself, potentially leading to failures during recording or replay. Without targeted enhancements for debugging, existing \recordreplay tools may struggle to meet the practical needs of developers when dealing with real-world software failures.}

In this paper, we conduct the first comprehensive study on the ability of \recordreplay tools to record and reliably replay non-crashing bugs and crashing bugs (referred to as \emph{failures}, and \emph{crashes}, respectively). In particular, we focus on buggy scenarios but also include common user scenarios~(referred to as just \emph{scenarios}) for comparison. To enhance the generalizability of our findings, we make use of two popular datasets and four state-of-the-art \recordreplay tools.
We find that the four \recordreplay tools exhibit substantially higher failure rates in buggy scenarios, with \perfailure{} for failures and \percrash{} for crashes, compared to a much lower failure rate of \perscenario{} for user scenarios.

In addition to reproducing bugs from manually produced traces, \recordreplay tools could also play a substantial role in reproducing bugs detected by automated input generation (\aig) techniques. After a crash is detected using \aig tools, developers and testers often rely on the stack trace to locate the root cause of the crash, rather than depending on the generated input traces.
One reason is the difficulty in reproducing a bug after the bug was detected by \aig tools (e.g., reproducing a crash after the \aig tool ran for 1 hour). 
Although, some \aig tools offer built-in functionality to assist with bug reproduction, this functionality is often inadequate.
For instance, Monkey~\cite{AndroidMonkey}, a popular \aig tool, accepts the $-s \langle seed \rangle$ in its command options to help reproduce bugs. By using the same seed, the sequence generated by Monkey should be consistent. However, the effectiveness of setting such seeds is often disappointing, e.g., in our experiments, we find that up to 74\% of cases cannot be reproduced when the same seed is used. 
Two potential ways to improve reproduction are to (1) improve \aig tools to reproduce crashes, and (2) use \aig tools with \recordreplay tools.

To examine the potential for reproducing crashes from \aig tools, our study involves the following investigations.
For (1), we aim to perform simple changes to \aig tools to understand whether such changes can help \aig tools reproduce crashes. Considering that \aig tools normally apply short action intervals (e.g., Monkey and \ape default use 200 ms), we conducted comparative experiments to measure the impact of different action intervals. The results indicate that even with increased action intervals, there is no substantial improvement in the outcomes (e.g., up to 70\% of cases for Monkey still failed). 
Although the original intent of \recordreplay tools are not meant to be combined with \aig tools, realizing the inadequacies of \aig tools themselves, we also explore (2), the use of \recordreplay tools with \aig tools to understand what changes are needed for \recordreplay tools to help reproduce crashes from \aig tools. Our findings show that even though (1) and (2) had similar results in reproducing \aig tool bugs, the limitations between them are different. 
By exploring the limitations in these two approaches, we aim to inspire and guide future work to improve and develop \aig tools and \recordreplay tools with better performance. Based on our observations, the direct improvements for (1) that future work can consider involve logging all actions on the fly by \aig tools and applying trace-reduction techniques to remove actions that are affecting the reproducibility of crashes. For (2), the primary obstacle to integrating \aig tools with \recordreplay tools is the compatibility issue with Android SDK tools.
\Space{
Specifically, our study investigates the following research questions (RQs):

\noindent\textbf{RQ$_1$}: How is the performance of existing trace-based \recordreplay tools to reproduce bugs from bug reports?

\noindent\textbf{RQ$_2$}: What are the root causes of unsuccessful cases in reproducing bugs from bug reports?

\noindent\textbf{RQ$_3$}: How is the performance of existing trace-based \recordreplay tools to reproduce bugs detected by \aig tools?

\noindent\textbf{RQ$_4$}: What are the root causes of unsuccessful cases in reproducing bugs detected by \aig tools?
}
The contributions of this paper are as follows:
\begin{itemize}
    \item{We present the first comprehensive study on manual recording and automated replaying of many common user scenarios, non-crashing failures, and crashing bugs for four popular \recordreplay tools. Table~\ref{tab:study:compare} provides an overview of the use cases we study, compares our study with prior work, and highlights the key knowledge gaps that our study aims to fill.}
    \item{We conduct a study investigating the extent to which \aig tools reliably replay discovered crashes.}
    \item{We highlight the limitations in current \recordreplay techniques and motivate future work to improve their reliability and consistency in different kinds of real-world use cases.}
    \item{We open-source our dataset~\cite{appendix} to aid future research.}
\end{itemize}

\section{Study Setup}
\label{sec:approach}

We study the potential for developers to use\Space{ trace-based} \recordreplay tools to manually record and automatically replay traces under three types of use cases: scenarios, failures, and crashes of popular Android apps.
Specifically, our study investigates the following research questions (RQs):

\noindent\textbf{RQ$_1$}: What is the performance of existing trace-based \recordreplay tools in reproducing bugs from bug reports?

\noindent\textbf{RQ$_2$}: What are the root causes of unsuccessful cases in reproducing bugs from bug reports?

\noindent\textbf{RQ$_3$}: What is the performance of existing trace-based \recordreplay tools in reproducing bugs detected by \aig tools?

\noindent\textbf{RQ$_4$}: What are the root causes of the unsuccessful cases in reproducing bugs detected by \aig tools?

\begin{figure*}[t]
    \centering
    \includegraphics[width=0.68\linewidth]{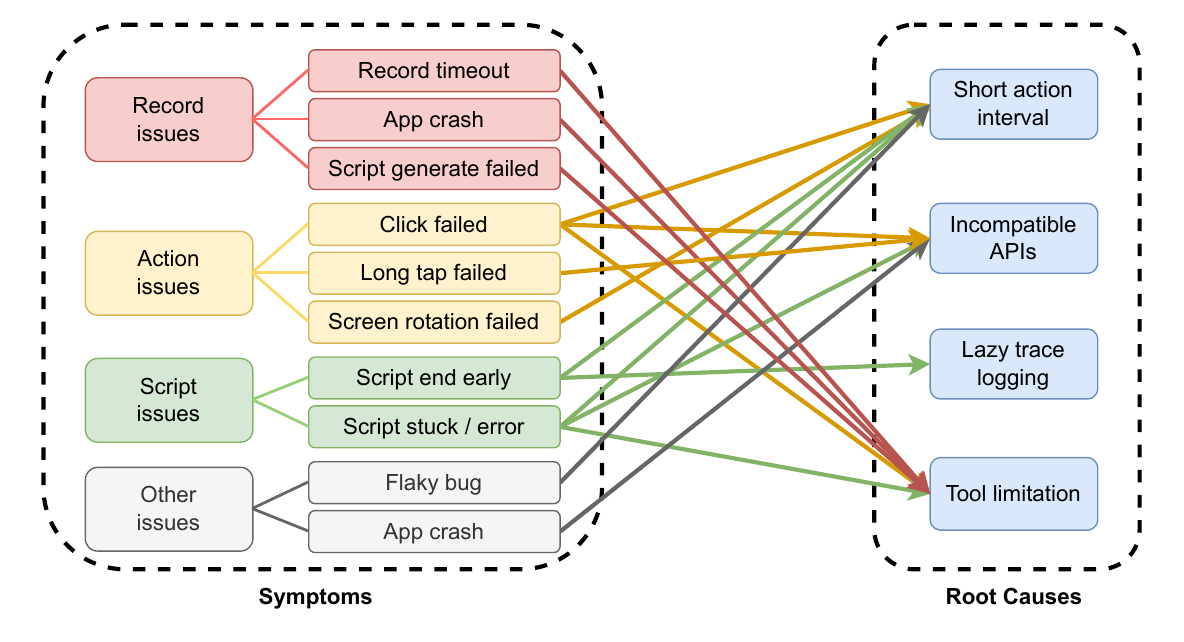}
    \caption{An overview of the relationship of issue symptoms and root causes}
 \label{fig:aig:rr_overview}
\end{figure*}

Our RQs aim to assess \recordreplay tools in two aspects. 
The first aspect (RQ$_1$ and RQ$_2$) is the ability to record and replay a given scenario, failure, or crash from bug reports. 
In this section, we describe this first aspect's experimental setup in detail. 
Figure~\ref{fig:aig:rr_overview} provides an overview of our analysis of all issues in reproduction from bug reports, illustrating the relationship between issue symptoms and their root causes. 
The second aspect (RQ$_3$ and RQ$_4$) is the ability of \recordreplay tools to record and replay crashes detected by \aig tools during their exploration. The experimental setup of this second aspect is described in Section~\ref{sec:rq3}.

\subsection{Methodology}
\label{sec:methodology}

For our study, we aim to record and replay scenarios, failures, and crashes. Our general methodology is as follows:
\begin{itemize}
    \item \textbf{Step 1}: \emph{Encounter} scenario, failure, or crash without any \recordreplay tool. For scenarios, we follow the designed steps to manually reproduce. For failures and crashes, we manually detect them following the steps outlined in bug reports.
    \item \textbf{Step 2}: \emph{Record} the scenario, failure, or crash with \recordreplay tools, i.e., we attempt to record the scenarios, failures, and crashes encountered in Step 1.
    \item \textbf{Step 3}: \emph{Replay} recorded scenarios, failures, and crashes\Space{, i.e., we attempt to replay the recorded scenarios, failures, and crashes} from Step 2.
\end{itemize}
The goal of Step 1 is to ensure that all scenarios, failures, and crashes used in our dataset are reproducible without \recordreplay tools.
The goal of Step 2 is to then evaluate how many use cases can each \recordreplay tool record.
Lastly, the goal of Step 3 is to evaluate the replayability of the use cases recorded by the \recordreplay tools.
The use cases that are not always successfully recorded and replayed in Steps 2 and 3 are then investigated to determine how the \recordreplay tools can be improved.
For Steps 1 and 2 of our study, we attempt to detect and record each scenario, failure, or crash three times (e.g., we follow the steps in a given bug report up to three times to record a trace demonstrating the corresponding failure or crash).
For Step 3, we attempt to replay each use case five times. We increased the replay times due to the observation of flakiness, i.e., the same trace exhibited inconsistent results across replays. 
To reduce the impact of flakiness on our replayability results, we increased the number of replays to five.

\subsection{Context}
\label{sec:context}

For our study, we use a set of scenarios and a set of detectable bugs for Android mobile testing. 
Our study aims to understand how effective and reliable these tools are in successfully recording and replaying various bugs that a developer might get from testers or end-users. 
The goal is to understand to what extent a developer can rely on \recordreplay tools.

\subsubsection{\textbf{Tools}}
\newcommand{\numRRTools}{11\xspace}
For this study, we surveyed prior papers on the topic of \recordreplay for Android and attempted to include recent state-of-the-art tools.
We found \numRRTools\Space{ trace-based} \recordreplay tools: Appium~\cite{appium}, Culebra~\cite{culebra}, GIFDroid~\cite{gifdroid}, Monkeyrunner~\cite{monkeyrunner}, Mosaic~\cite{mosaic}, Robotium recorder~\cite{Robotium}, \reran~\cite{reran}, \replaykit~\cite{replaykit}, \sara~\cite{sara}, VALERA~\cite{VALERA}, and V2S~\cite{V2S}.

Most of the tools we found are mentioned in a prior study~\cite{lam17:record} on \recordreplay tools, which found that the tools were often inadequate for developers to use. 
Taking into account time and labor costs, as well as the diversity and advancement of tools, we selected the following four tools. Four participants, including three doctoral students and one master's student, each used one of these tools to conduct all related experiments.
(a)~\emph{\replaykit} is a command-line tool of Appetizer. It can record, replay, and mirror touchscreen events for Android. Similar to \reran, \replaykit also relies on Android SDK tools to record and replay events.
(b)~\emph{\reran} is a lightweight desktop tool that records UI traces based on coordinates and states. It is one of the very first \recordreplay tools for research purposes. It works in the kernel layer, capturing and replaying events via Android SDK tools \textit{getevent} and \textit{sendevent}. 
(c)~\emph{\sara} is a recent state-of-the-art \recordreplay tool that specifically aims to mitigate some of the limitations found in other tools. As the SARA tool is published after the most recent study~\cite{lam17:record} on \recordreplay tools, it is particularly unclear how \sara compares to other \recordreplay tools in terms of recordability and replayability. \sara uses dynamic instrumentation to record and replay diverse kinds of inputs in mobile devices, and the self-replay mechanism can address the problem of recording motion events based on widgets. 
(d)~\emph{V2S} is the first Android tool that automates the analysis of video-based mobile development. V2S processes the screen recordings of Android app usages by applying constraints on frame size,  each video should be recorded with at least 30 FPS, and users should enable the ``Show Touches'' option in developer mode. Then, V2S detects and classifies actions using deep learning techniques and translates the actions into a replayable trace for a given target device.

\subsubsection{\textbf{Dataset}} Our dataset includes 34 scenarios, 90 failures and 31 crashes in total.

\noindent\textbf{Scenarios.} To obtain scenarios, we use apps in the Themis dataset~\cite{Su2021WS}.
The Themis dataset consists of 52 crashes from 20 popular apps. 
For each app, we follow the method in prior work~\cite{sara} to design two scenarios according to the description and images of its GitHub and Google Play homepage. The selected scenarios include common user activities like adding/editing/searching/selecting contents (e.g., notes \& forms), navigating menu items, file operations (e.g., explore directories \& copy-paste files), and multimedia operations (e.g., capture \& save pictures). 
For example, for the Frost app~\cite{frost}, a third-party Facebook app for Android, one of the scenarios we designed is to create a post. 
For the FirefoxLite app~\cite{firefox}, one of the scenarios we designed is to do a keyword search on the homepage and jump to one of the searched web pages. 
Three apps were removed from our list because the main functionalities were no longer worked to keep the reported crash existing.

\noindent\textbf{Failures.} We use the AndroR2 dataset~\cite{androR2dataset}, which is a collection of Android bug reports mined from popular Android apps that are hosted on GitHub and cross-listed on Google Play. 
Each bug report includes reproduction steps with the app version, Android OS version, screenshots, videos, an APK file, and the GitHub link to the bug report. We select all failures from the dataset and remove the ones that are detected on only Android versions 10 and 11 since one of the \recordreplay tools (\reran) cannot replay on such versions. 

After this filtering process, we are left with 116 of 180 bugs to analyze for our study. 
For Step 1, for each bug report present in AndroR2, four authors (one for each \recordreplay tool) discuss and agree on guidelines to be followed in recording each of the bugs, then each author attempts to follow the reproduction steps given in the bug report and evaluate whether they could reproduce the described bug manually. 
Each author creates a virtual Android device (emulator) that matches the specified Android OS version stated in each bug report and makes three attempts to reproduce the bug.
In the end, we can manually reproduce 90 of 116 failures. 
26 bugs are removed from our dataset due to the following three reasons: (1) the reproduction steps were unclear in the bug report, (2) the bug is no longer present in the app (e.g., the server of the app has disabled certain functionality), and (3) the bug is blocked by another bug in the app.

\noindent\textbf{Crashes.} We use 52 verified crashes from the Themis dataset. 
We repeat the Step 1 process of the failure dataset for crashes and obtain 31 usable crashes in the end. 
21 crashes are removed from our study due to (1)~the crash is not reproducible, (2)~the app cannot be installed properly on our devices, and (3)~the reproduction steps were unclear in the bug report.

\section{RQ$_1$: Reproduction from bug reports}
\label{sec:results}

\begin{table*}[hbpt]
  \centering
  \caption{Breakdown of unsuccessful recording \& replaying cases. The percentages of unsuccessful cases for user scenarios, failures, and crashes are also shown.}

\begin{tabular}{|lllrrrrr|}
\hline
\multicolumn{1}{|l|}{\textbf{Steps}}                                           & \multicolumn{1}{l|}{\textbf{Categories}}                                     & \multicolumn{1}{l|}{\textbf{Subcategories}}                         & \multicolumn{1}{c|}{\textbf{RERAN}}             & \multicolumn{1}{c|}{\textbf{ReplayKit}}         & \multicolumn{1}{c|}{\textbf{SARA}}              & \multicolumn{1}{c|}{\textbf{V2S}}               & \multicolumn{1}{c|}{\textbf{Count}} \\ \hline
&&&&&&&\\[-0.5em]
\multicolumn{8}{|l|}{USER SCENARIOS}                                                                                                                                                                                                                                                                                                                                                                                                                                             \\ \hline
\rowcolor[HTML]{D9EAD3} 
\multicolumn{1}{|l|}{\cellcolor[HTML]{D9EAD3}}                                & \multicolumn{2}{l|}{\cellcolor[HTML]{D9EAD3}\sara Frida error}                                                                                           & \multicolumn{1}{r|}{\cellcolor[HTML]{D9EAD3}-}  & \multicolumn{1}{r|}{\cellcolor[HTML]{D9EAD3}-}  & \multicolumn{1}{r|}{\cellcolor[HTML]{D9EAD3}6}  & \multicolumn{1}{r|}{\cellcolor[HTML]{D9EAD3}-}  & 6                                   \\ \cline{2-8} 
\rowcolor[HTML]{D9EAD3} 
\multicolumn{1}{|l|}{\multirow{-2}{*}{\cellcolor[HTML]{D9EAD3}Record}} & \multicolumn{2}{l|}{\cellcolor[HTML]{D9EAD3}\textbf{Count}}                                                                                                 & \multicolumn{1}{r|}{\cellcolor[HTML]{D9EAD3}\textbf{-}}  & \multicolumn{1}{r|}{\cellcolor[HTML]{D9EAD3}\textbf{-}}  & \multicolumn{1}{r|}{\cellcolor[HTML]{D9EAD3}\textbf{6}}  & \multicolumn{1}{r|}{\cellcolor[HTML]{D9EAD3}\textbf{-}}  & \textbf{6}                                   \\ \hline
\rowcolor[HTML]{FFFFC7} 
\multicolumn{1}{|l|}{\cellcolor[HTML]{FFFFC7}}                                & \multicolumn{1}{l|}{\cellcolor[HTML]{FFFFC7}Action issues}                    & \multicolumn{1}{l|}{\cellcolor[HTML]{FFFFC7}Click failed}           & \multicolumn{1}{r|}{\cellcolor[HTML]{FFFFC7}2}  & \multicolumn{1}{r|}{\cellcolor[HTML]{FFFFC7}-}  & \multicolumn{1}{r|}{\cellcolor[HTML]{FFFFC7}2}  & \multicolumn{1}{r|}{\cellcolor[HTML]{FFFFC7}8}  & 12                                  \\ \cline{2-8} 
\rowcolor[HTML]{FFFFC7} 
\multicolumn{1}{|l|}{\cellcolor[HTML]{FFFFC7}}                                & \multicolumn{1}{l|}{\cellcolor[HTML]{FFFFC7}}                                & \multicolumn{1}{l|}{\cellcolor[HTML]{FFFFC7}Script stuck/error}     & \multicolumn{1}{r|}{\cellcolor[HTML]{FFFFC7}-}  & \multicolumn{1}{r|}{\cellcolor[HTML]{FFFFC7}-}  & \multicolumn{1}{r|}{\cellcolor[HTML]{FFFFC7}4}  & \multicolumn{1}{r|}{\cellcolor[HTML]{FFFFC7}-}  & 4                                   \\ \cline{3-8} 
\rowcolor[HTML]{FFFFC7} 
\multicolumn{1}{|l|}{\cellcolor[HTML]{FFFFC7}}                                & \multicolumn{1}{l|}{\multirow{-2}{*}{\cellcolor[HTML]{FFFFC7}Script issues}}  & \multicolumn{1}{l|}{\cellcolor[HTML]{FFFFC7}Script ending early}    & \multicolumn{1}{r|}{\cellcolor[HTML]{FFFFC7}-}  & \multicolumn{1}{r|}{\cellcolor[HTML]{FFFFC7}-}  & \multicolumn{1}{r|}{\cellcolor[HTML]{FFFFC7}-}  & \multicolumn{1}{r|}{\cellcolor[HTML]{FFFFC7}1}  & 1                                   \\ \cline{2-8} 
\rowcolor[HTML]{FFFFC7} 
\multicolumn{1}{|l|}{\multirow{-4}{*}{\cellcolor[HTML]{FFFFC7}Replay}} & \multicolumn{2}{l|}{\cellcolor[HTML]{FFFFC7}\textbf{Count}}                                                                                                 & \multicolumn{1}{r|}{\cellcolor[HTML]{FFFFC7}\textbf{2}}  & \multicolumn{1}{r|}{\cellcolor[HTML]{FFFFC7}\textbf{-}}  & \multicolumn{1}{r|}{\cellcolor[HTML]{FFFFC7}\textbf{6}}  & \multicolumn{1}{r|}{\cellcolor[HTML]{FFFFC7}\textbf{9}}  & \textbf{17}                                  \\ \hline
\multicolumn{3}{|l|}{\textbf{Total (unsuccessful cases / \# of user scenarios (34))}}                                                                                                                                                                                               & \multicolumn{1}{r|}{\textbf{2 (6\%)}}        & \multicolumn{1}{r|}{\textbf{0 (0\%)}}        & \multicolumn{1}{r|}{\textbf{12 (35\%)}}      & \multicolumn{1}{r|}{\textbf{9 (27\%)}}       & \textbf{23 (17\%)}               \\ \hline

&&&&&&&\\[-0.5em]
\multicolumn{8}{|l|}{FAILURES}                                                                                                                                                                                                                                                                                                                                                                                                                                                   \\ \hline
\rowcolor[HTML]{D9EAD3} 
\multicolumn{1}{|l|}{\cellcolor[HTML]{D9EAD3}}                                & \multicolumn{2}{l|}{\cellcolor[HTML]{D9EAD3}\sara Frida error}                                                                                           & \multicolumn{1}{r|}{\cellcolor[HTML]{D9EAD3}-}  & \multicolumn{1}{r|}{\cellcolor[HTML]{D9EAD3}-}  & \multicolumn{1}{r|}{\cellcolor[HTML]{D9EAD3}21} & \multicolumn{1}{r|}{\cellcolor[HTML]{D9EAD3}-}  & 21                                  \\ \cline{2-8}
\rowcolor[HTML]{D9EAD3} 
\multicolumn{1}{|l|}{\cellcolor[HTML]{D9EAD3}}                                & \multicolumn{2}{l|}{\cellcolor[HTML]{D9EAD3}Generate script failed}                                                                                & \multicolumn{1}{r|}{\cellcolor[HTML]{D9EAD3}-}  & \multicolumn{1}{r|}{\cellcolor[HTML]{D9EAD3}-}  & \multicolumn{1}{r|}{\cellcolor[HTML]{D9EAD3}-}  & \multicolumn{1}{r|}{\cellcolor[HTML]{D9EAD3}9}  & 9                                   \\ \cline{2-8}
\rowcolor[HTML]{D9EAD3} 
\multicolumn{1}{|l|}{\cellcolor[HTML]{D9EAD3}}                                & \multicolumn{2}{l|}{\cellcolor[HTML]{D9EAD3}App crash}                                                                                             & \multicolumn{1}{r|}{\cellcolor[HTML]{D9EAD3}-}  & \multicolumn{1}{r|}{\cellcolor[HTML]{D9EAD3}-}  & \multicolumn{1}{r|}{\cellcolor[HTML]{D9EAD3}8}  & \multicolumn{1}{r|}{\cellcolor[HTML]{D9EAD3}-}  & 8                                   \\ \cline{2-8} 
\rowcolor[HTML]{D9EAD3} 
\multicolumn{1}{|l|}{\multirow{-4}{*}{\cellcolor[HTML]{D9EAD3}Record}} & \multicolumn{2}{l|}{\cellcolor[HTML]{D9EAD3}\textbf{Count}}                                                                                                 & \multicolumn{1}{r|}{\cellcolor[HTML]{D9EAD3}\textbf{-}}  & \multicolumn{1}{r|}{\cellcolor[HTML]{D9EAD3}\textbf{-}}  & \multicolumn{1}{r|}{\cellcolor[HTML]{D9EAD3}\textbf{29}} & \multicolumn{1}{r|}{\cellcolor[HTML]{D9EAD3}\textbf{9}}  & \textbf{38}                                  \\ \hline
\rowcolor[HTML]{FFFFC7} 
\multicolumn{1}{|l|}{\cellcolor[HTML]{FFFFC7}}                                & \multicolumn{1}{l|}{\cellcolor[HTML]{FFFFC7}}                                & \multicolumn{1}{l|}{\cellcolor[HTML]{FFFFC7}Click failed}           & \multicolumn{1}{r|}{\cellcolor[HTML]{FFFFC7}14} & \multicolumn{1}{r|}{\cellcolor[HTML]{FFFFC7}-}  & \multicolumn{1}{r|}{\cellcolor[HTML]{FFFFC7}15} & \multicolumn{1}{r|}{\cellcolor[HTML]{FFFFC7}24} & 53                                  \\ \cline{3-8} 
\rowcolor[HTML]{FFFFC7} 
\multicolumn{1}{|l|}{\cellcolor[HTML]{FFFFC7}}                                & \multicolumn{1}{l|}{\cellcolor[HTML]{FFFFC7}}                                & \multicolumn{1}{l|}{\cellcolor[HTML]{FFFFC7}Screen rotation failed} & \multicolumn{1}{r|}{\cellcolor[HTML]{FFFFC7}1}  & \multicolumn{1}{r|}{\cellcolor[HTML]{FFFFC7}5}  & \multicolumn{1}{r|}{\cellcolor[HTML]{FFFFC7}2}  & \multicolumn{1}{r|}{\cellcolor[HTML]{FFFFC7}2}  & 10                                  \\ \cline{3-8} 
\rowcolor[HTML]{FFFFC7} 
\multicolumn{1}{|l|}{\cellcolor[HTML]{FFFFC7}}                                & \multicolumn{1}{l|}{\multirow{-3}{*}{\cellcolor[HTML]{FFFFC7}Action issues}} & \multicolumn{1}{l|}{\cellcolor[HTML]{FFFFC7}Long tap failed}        & \multicolumn{1}{r|}{\cellcolor[HTML]{FFFFC7}-}  & \multicolumn{1}{r|}{\cellcolor[HTML]{FFFFC7}7}  & \multicolumn{1}{r|}{\cellcolor[HTML]{FFFFC7}-}  & \multicolumn{1}{r|}{\cellcolor[HTML]{FFFFC7}2}  & 9                                   \\ \cline{2-8} 
\rowcolor[HTML]{FFFFC7} 
\multicolumn{1}{|l|}{\cellcolor[HTML]{FFFFC7}}                                & \multicolumn{1}{l|}{\cellcolor[HTML]{FFFFC7}}                                & \multicolumn{1}{l|}{\cellcolor[HTML]{FFFFC7}Script ending early}    & \multicolumn{1}{r|}{\cellcolor[HTML]{FFFFC7}9}  & \multicolumn{1}{r|}{\cellcolor[HTML]{FFFFC7}0}  & \multicolumn{1}{r|}{\cellcolor[HTML]{FFFFC7}2}  & \multicolumn{1}{r|}{\cellcolor[HTML]{FFFFC7}0}  & 11                                  \\ \cline{3-8} 
\rowcolor[HTML]{FFFFC7} 
\multicolumn{1}{|l|}{\cellcolor[HTML]{FFFFC7}}                                & \multicolumn{1}{l|}{\multirow{-2}{*}{\cellcolor[HTML]{FFFFC7}Script issues}} & \multicolumn{1}{l|}{\cellcolor[HTML]{FFFFC7}Script stuck/error}     & \multicolumn{1}{r|}{\cellcolor[HTML]{FFFFC7}-}  & \multicolumn{1}{r|}{\cellcolor[HTML]{FFFFC7}-}  & \multicolumn{1}{r|}{\cellcolor[HTML]{FFFFC7}-}  & \multicolumn{1}{r|}{\cellcolor[HTML]{FFFFC7}7}  & 7                                   \\ \cline{2-8} 
\rowcolor[HTML]{FFFFC7} 
\multicolumn{1}{|l|}{\cellcolor[HTML]{FFFFC7}}                                & \multicolumn{1}{l|}{\cellcolor[HTML]{FFFFC7}}                                & \multicolumn{1}{l|}{\cellcolor[HTML]{FFFFC7}App crash}              & \multicolumn{1}{r|}{\cellcolor[HTML]{FFFFC7}-}  & \multicolumn{1}{r|}{\cellcolor[HTML]{FFFFC7}-}  & \multicolumn{1}{r|}{\cellcolor[HTML]{FFFFC7}5}  & \multicolumn{1}{r|}{\cellcolor[HTML]{FFFFC7}1}  & 6                                   \\ \cline{3-8} 
\rowcolor[HTML]{FFFFC7} 
\multicolumn{1}{|l|}{\cellcolor[HTML]{FFFFC7}}                                & \multicolumn{1}{l|}{\cellcolor[HTML]{FFFFC7}}                                & \multicolumn{1}{l|}{\cellcolor[HTML]{FFFFC7}Camera open failed}     & \multicolumn{1}{r|}{\cellcolor[HTML]{FFFFC7}-}  & \multicolumn{1}{r|}{\cellcolor[HTML]{FFFFC7}1}  & \multicolumn{1}{r|}{\cellcolor[HTML]{FFFFC7}-}  & \multicolumn{1}{r|}{\cellcolor[HTML]{FFFFC7}-}  & 1                                   \\ \cline{3-8} 
\rowcolor[HTML]{FFFFC7} 
\multicolumn{1}{|l|}{\cellcolor[HTML]{FFFFC7}}                                & \multicolumn{1}{l|}{\multirow{-3}{*}{\cellcolor[HTML]{FFFFC7}Other}}         & \multicolumn{1}{l|}{\cellcolor[HTML]{FFFFC7}Flaky bug}              & \multicolumn{1}{r|}{\cellcolor[HTML]{FFFFC7}1}  & \multicolumn{1}{r|}{\cellcolor[HTML]{FFFFC7}-}  & \multicolumn{1}{r|}{\cellcolor[HTML]{FFFFC7}1}  & \multicolumn{1}{r|}{\cellcolor[HTML]{FFFFC7}-}  & 2                                   \\ \cline{2-8} 
\rowcolor[HTML]{FFFFC7} 
\multicolumn{1}{|l|}{\multirow{-9}{*}{\cellcolor[HTML]{FFFFC7}Replay}} & \multicolumn{2}{l|}{\cellcolor[HTML]{FFFFC7}\textbf{Count}}                                                                                                 & \multicolumn{1}{r|}{\cellcolor[HTML]{FFFFC7}\textbf{25}} & \multicolumn{1}{r|}{\cellcolor[HTML]{FFFFC7}\textbf{13}} & \multicolumn{1}{r|}{\cellcolor[HTML]{FFFFC7}\textbf{25}} & \multicolumn{1}{r|}{\cellcolor[HTML]{FFFFC7}\textbf{36}} & \textbf{99}                                  \\ \hline
\multicolumn{3}{|l|}{\textbf{Total (unsuccessful cases / \# of failures (90))}}                                                                                                                                                                                               & \multicolumn{1}{r|}{\textbf{25 (28\%)}}      & \multicolumn{1}{r|}{\textbf{13 (14\%)}}      & \multicolumn{1}{r|}{\textbf{54 (60\%)}}      & \multicolumn{1}{r|}{\textbf{45 (50\%)}}      & \textbf{137 (38\%)}              \\ \hline

&&&&&&&\\[-0.5em]
\multicolumn{8}{|l|}{CRASHES}                                                                                                                                                                                                                                                                                                                                                                                                                                                    \\ \hline
\rowcolor[HTML]{D9EAD3} 
\multicolumn{1}{|l|}{\cellcolor[HTML]{D9EAD3}}                                & \multicolumn{2}{l|}{\cellcolor[HTML]{D9EAD3}App crash}                                                                                             & \multicolumn{1}{r|}{\cellcolor[HTML]{D9EAD3}-}  & \multicolumn{1}{r|}{\cellcolor[HTML]{D9EAD3}-}  & \multicolumn{1}{r|}{\cellcolor[HTML]{D9EAD3}4}  & \multicolumn{1}{r|}{\cellcolor[HTML]{D9EAD3}-}  & 4                                   \\ \cline{2-8} 
\rowcolor[HTML]{D9EAD3} 
\multicolumn{1}{|l|}{\cellcolor[HTML]{D9EAD3}}                                & \multicolumn{2}{l|}{\cellcolor[HTML]{D9EAD3}Track action failed}                                                                                   & \multicolumn{1}{r|}{\cellcolor[HTML]{D9EAD3}-}  & \multicolumn{1}{r|}{\cellcolor[HTML]{D9EAD3}-}  & \multicolumn{1}{r|}{\cellcolor[HTML]{D9EAD3}2}  & \multicolumn{1}{r|}{\cellcolor[HTML]{D9EAD3}-}  & 2                                   \\ \cline{2-8} 
\rowcolor[HTML]{D9EAD3} 
\multicolumn{1}{|l|}{\multirow{-3}{*}{\cellcolor[HTML]{D9EAD3}Record}} & \multicolumn{2}{l|}{\cellcolor[HTML]{D9EAD3}\textbf{Count}}                                                                                                 & \multicolumn{1}{r|}{\cellcolor[HTML]{D9EAD3}\textbf{-}}  & \multicolumn{1}{r|}{\cellcolor[HTML]{D9EAD3}\textbf{-}}  & \multicolumn{1}{r|}{\cellcolor[HTML]{D9EAD3}\textbf{6}}  & \multicolumn{1}{r|}{\cellcolor[HTML]{D9EAD3}\textbf{-}}  & \textbf{6}                                  \\ \hline
\rowcolor[HTML]{FFFFC7} 
\multicolumn{1}{|l|}{\cellcolor[HTML]{FFFFC7}}                                & \multicolumn{1}{l|}{\cellcolor[HTML]{FFFFC7}}                                & \multicolumn{1}{l|}{\cellcolor[HTML]{FFFFC7}Click failed}           & \multicolumn{1}{r|}{\cellcolor[HTML]{FFFFC7}6}  & \multicolumn{1}{r|}{\cellcolor[HTML]{FFFFC7}1}  & \multicolumn{1}{r|}{\cellcolor[HTML]{FFFFC7}6}  & \multicolumn{1}{r|}{\cellcolor[HTML]{FFFFC7}14} & 27                                  \\ \cline{3-8} 
\rowcolor[HTML]{FFFFC7} 
\multicolumn{1}{|l|}{\cellcolor[HTML]{FFFFC7}}                                & \multicolumn{1}{l|}{\cellcolor[HTML]{FFFFC7}}                                & \multicolumn{1}{l|}{\cellcolor[HTML]{FFFFC7}Long  tap failed}       & \multicolumn{1}{r|}{\cellcolor[HTML]{FFFFC7}-}  & \multicolumn{1}{r|}{\cellcolor[HTML]{FFFFC7}6}  & \multicolumn{1}{r|}{\cellcolor[HTML]{FFFFC7}-}  & \multicolumn{1}{r|}{\cellcolor[HTML]{FFFFC7}1}  & 7                                   \\ \cline{3-8} 
\rowcolor[HTML]{FFFFC7} 
\multicolumn{1}{|l|}{\cellcolor[HTML]{FFFFC7}}                                & \multicolumn{1}{l|}{\multirow{-3}{*}{\cellcolor[HTML]{FFFFC7}Action issues}} & \multicolumn{1}{l|}{\cellcolor[HTML]{FFFFC7}Screen rotation failed} & \multicolumn{1}{r|}{\cellcolor[HTML]{FFFFC7}-}  & \multicolumn{1}{r|}{\cellcolor[HTML]{FFFFC7}1}  & \multicolumn{1}{r|}{\cellcolor[HTML]{FFFFC7}-}  & \multicolumn{1}{r|}{\cellcolor[HTML]{FFFFC7}1}  & 2                                   \\ \cline{2-8} 
\rowcolor[HTML]{FFFFC7} 
\multicolumn{1}{|l|}{\cellcolor[HTML]{FFFFC7}}                                & \multicolumn{1}{l|}{\cellcolor[HTML]{FFFFC7}}                                & \multicolumn{1}{l|}{\cellcolor[HTML]{FFFFC7}Script ending early}    & \multicolumn{1}{r|}{\cellcolor[HTML]{FFFFC7}-}  & \multicolumn{1}{r|}{\cellcolor[HTML]{FFFFC7}1}  & \multicolumn{1}{r|}{\cellcolor[HTML]{FFFFC7}-}  & \multicolumn{1}{r|}{\cellcolor[HTML]{FFFFC7}-}  & 1                                   \\ \cline{3-8} 
\rowcolor[HTML]{FFFFC7} 
\multicolumn{1}{|l|}{\cellcolor[HTML]{FFFFC7}}                                & \multicolumn{1}{l|}{\multirow{-2}{*}{\cellcolor[HTML]{FFFFC7}Script issues}} & \multicolumn{1}{l|}{\cellcolor[HTML]{FFFFC7}Script stuck/error}     & \multicolumn{1}{r|}{\cellcolor[HTML]{FFFFC7}-}  & \multicolumn{1}{r|}{\cellcolor[HTML]{FFFFC7}-}  & \multicolumn{1}{r|}{\cellcolor[HTML]{FFFFC7}1}  & \multicolumn{1}{r|}{\cellcolor[HTML]{FFFFC7}-}  & 1                                   \\ \cline{2-8} 
\rowcolor[HTML]{FFFFC7} 
\multicolumn{1}{|l|}{\cellcolor[HTML]{FFFFC7}}                                & \multicolumn{1}{l|}{\cellcolor[HTML]{FFFFC7}}                                & \multicolumn{1}{l|}{\cellcolor[HTML]{FFFFC7}App crash}              & \multicolumn{1}{r|}{\cellcolor[HTML]{FFFFC7}5}  & \multicolumn{1}{r|}{\cellcolor[HTML]{FFFFC7}-}  & \multicolumn{1}{r|}{\cellcolor[HTML]{FFFFC7}4}  & \multicolumn{1}{r|}{\cellcolor[HTML]{FFFFC7}-}  & 9                                   \\ \cline{3-8} 
\rowcolor[HTML]{FFFFC7} 
\multicolumn{1}{|l|}{\cellcolor[HTML]{FFFFC7}}                                & \multicolumn{1}{l|}{\multirow{-2}{*}{\cellcolor[HTML]{FFFFC7}Other}}         & \multicolumn{1}{l|}{\cellcolor[HTML]{FFFFC7}Flaky bug}              & \multicolumn{1}{r|}{\cellcolor[HTML]{FFFFC7}-}  & \multicolumn{1}{r|}{\cellcolor[HTML]{FFFFC7}-}  & \multicolumn{1}{r|}{\cellcolor[HTML]{FFFFC7}1}  & \multicolumn{1}{r|}{\cellcolor[HTML]{FFFFC7}-}  & 1                                   \\ \cline{2-8} 
\rowcolor[HTML]{FFFFC7} 
\multicolumn{1}{|l|}{\multirow{-8}{*}{\cellcolor[HTML]{FFFFC7}Replay}} & \multicolumn{2}{l|}{\cellcolor[HTML]{FFFFC7}\textbf{Count}}                                                                                                 & \multicolumn{1}{r|}{\cellcolor[HTML]{FFFFC7}\textbf{11}} & \multicolumn{1}{r|}{\cellcolor[HTML]{FFFFC7}\textbf{9}}  & \multicolumn{1}{r|}{\cellcolor[HTML]{FFFFC7}\textbf{12}} & \multicolumn{1}{r|}{\cellcolor[HTML]{FFFFC7}\textbf{16}} & \textbf{48}                                  \\ \hline
\multicolumn{3}{|l|}{\textbf{Total (unsuccessful cases / \# of crashes (31))}}                                                                                                                                                                                               & \multicolumn{1}{r|}{\textbf{11 (36\%)}}      & \multicolumn{1}{r|}{\textbf{9 (29\%)}}       & \multicolumn{1}{r|}{\textbf{18 (58\%)}}      & \multicolumn{1}{r|}{\textbf{16 (52\%)}}      & \textbf{54 (44\%)}               \\ \hline
\end{tabular}
\label{tab:manual_table3}
\vspace{-1.5em}
\end{table*}
Table~\ref{tab:manual_table3} shows the number of cases each \recordreplay tool failed to record or replay. 
We use four \recordreplay tools to record all use cases in our dataset, which includes 34 scenarios, 90 failures, and 31 crashes, and then replay the successfully recorded use cases with each of the \recordreplay tools five times. 

\textbf{Unsuccessful cases in recording phase. }Among the four tools, only \reran and \replaykit could finish recording all cases. V2S did not generate recordings from videos for nine cases. 
Most unsuccessful recording cases come from \sara. Six scenarios and 21 failures cannot be recorded by \sara due to Frida error. Frida~\cite{frida} is a dynamic code instrumentation toolkit, which is used by \sara to instrument apps and monitor events. 
Beyond Frida errors, \sara cannot finish recording eight failures and four crashes because the app crashes as soon as recording starts.

\textbf{Unsuccessful cases in replaying phase. }The symptoms revealed in the replay process can be divided into two main categories, action issues and script issues.
Action issues refer to the situation in which a certain key action in the sequence is not successfully executed during the replay process, resulting in the deviation of all subsequent actions and replay failure. The key action could be a single click, long press and screen rotation, etc. 
Script issues are related to the quality of the generated replaying scripts. If the script is incomplete, e.g., missing a few actions at the end of the sequence, then the replay will end early before reproducing the targeted scenario. In other cases, the incorrect script could freeze the process or raise exceptions during execution.
Besides the two main categories, other symptoms include unexpected app crashes, failed to open the camera, etc.

\textbf{Performance for different types of use cases. }As illustrated in Table~\ref{tab:manual_table3}, we observe a substantial drop in performance for \recordreplay tools when they are used to record and replay failures and crashes compared to scenarios. 
For example, RERAN has a substantially higher percentage of unsuccessful cases for failures (28\%) and crashes (36\%) than for scenarios  (6\%). 
We observe the same trend for ReplayKit, SARA, and V2S. 
To better understand this substantial performance drop in replaying failures and crashes, we further investigate V2S's replay cases. 
Specifically, when we calculate the duration of videos recorded for V2S replay, we find that 
the average duration of scenarios is around 20 seconds, while the average duration of failures and crashes is around 40 seconds. 
Our observations suggest that scenarios are more trivial use cases that tend to represent the happy path of an app whereas failures and crashes tend to exhibit more complex corner cases.

\section{RQ$_2$: Root causes of issues in reproduction from bug reports}

After analyzing the symptoms of replay issues, we further investigate the root causes, which can be divided into four main categories (Figure~\ref{fig:aig:rr_overview}). Different symptoms may lead to the same root cause, and different root causes may also have the same symptoms. More details are in our appendix~\cite{appendix}.

\subsection{Short action interval}
Click failed is the most frequent issue that appears in our study. This category includes two situations: (1)~missed action during the recording process, i.e., the relevant click action was not successfully recorded; (2)~the click action did not apply to the correct component. 
Short action interval is an important root cause of click failed issues. 
If actions are entered at a fast frequency, \recordreplay tools are more likely to miss some actions and result in an unsuccessful replay. 
Furthermore, if the sequence involves page switching or network loading, a short action interval can easily cause actions to be executed on a screen that has not finished loading, causing subsequent actions to be invalid and the replay to be unsuccessful. 

Note that in our study results \reran and \sara have many more click failed cases than \replaykit. 
As mentioned before, in our study, we have four participants, each handling one \recordreplay tool and implementing all experiments of a particular tool. 
We conducted action interval surveys among the four participants, and it was found that the action interval used by \reran and \sara participants was about 0.5 seconds per action (s/a), while the participants using \replaykit and V2S had an action interval of 2 s/a. 
To confirm the impact of the action interval on our experimental results, we randomly selected three cases from the Click failed cases of \reran and \sara, and re-tested them using an action interval of 2 s/a. 
The results found that \reran could successfully replay all three cases. \sara successfully replayed two cases. 
For the last use case, \sara failed to replay due to a screen rotation issue. However, the action that caused the click to fail was correctly recorded and replayed this time. In addition, we also conducted similar experiments on \replaykit, where we randomly selected 3 successful cases and re-tested the tool with a short action interval of 0.5 s/a. 
We successfully reproduced one case that has a short action sequence and were unsuccessful in reproducing two other cases that have longer action sequences.

\textbf{Suggestions: }(1)~Longer action intervals that more closely resemble real-world user interactions and the avoidance of rapid-fire action sequences are more likely to result in successfully recorded traces and a higher likelihood to successfully replay the recorded traces; (2)~Introducing mechanisms\Space{ or error handlers in the replay process} to retry failed actions or ensuring that async actions and background processes are properly synchronized in the recording and replaying phases can help deal with load time related problems.\Space{ Implementing wait conditions can help ensure that actions are executed in the correct order and timing.}

\subsection{Incompatible APIs} 

Many unsuccessful replay cases are caused by incompatible API calls, such as screen rotation failed, long tap failed, camera open failed, and click failed cases. For instance, \replaykit cannot correctly record and replay long tap actions or open the camera interface. \sara and \replaykit often fail to perform screen rotation actions. \sara is also not yet able to record and replay actions involving the \textit{android.text.Editable} interface, which provides APIs to process input strings\Space{ because Frida's underlying logic does not support it}.

\textbf{Suggestions: }(1) Better API support offers richer functionality, which can help resolve issues related to complex actions, such as long taps and screen rotations, and lead to more effective recording and replaying; (2) Android's accessibility services can be leveraged to simulate touch interactions and screen rotations programmatically. Creating custom accessibility service modules can help \recordreplay tools support more functionalities, such as long tap and screen rotation.

\subsection{Lazy trace logging} 

When using the Android SDK \textit{getevent} tool to record actions, the actions are logged on the device storage. However, this process employs a lazy logging mechanism, meaning that the trace log will be updated only after a certain number of events have been accumulated. As a result, the final portion of actions may be missed if the number of events recorded is insufficient to trigger the logging process, which leads to issues where the scripts end early in the replaying phase.

\OurComment{
When using the Android SDK \textit{getevent} tool to record actions, the actions are logged on the device storage. However, this logging process employs a lazy logging mechanism, which means that the trace log will be updated only after a certain number of events have been accumulated. If the number of events recorded is insufficient to trigger the logging process, then the logging of actions may be skipped and result in empty scripts.
}

\textbf{Suggestions: }(1) Introducing an event buffering mechanism to temporarily store actions before updating the trace log; (2) Modifying the threshold for triggering the logging process to a lower value; (3) Adding additional actions after the recording to ensure that all targeted actions are recorded.

\subsection{Tool limitations\Space{: Trade-off between resource and information}}

\subsubsection{Trade-off between resource and information}
\label{sec:sara-io}
Frida error frequently happens during the recording phase. Such issues arise mainly due to \sara using the instrumentation tool Frida~\cite{frida} to instrument the device and monitor events. To support the cross-device replay feature, \sara records all relevant widgets and state information through the Frida tool during the recording process to facilitate subsequent filtering of relevant widgets for cross-device replay. This results in a substantial amount of I/O operations during the recording process, and introduces additional overhead which leads to delays in recording actions, causing the process to stall and eventually triggering a timeout error. Moreover, \sara also has many unsuccessful cases related to script stuck/error or click failures that are also related to I/O operations. 
The large influx of I/O operations from \sara recording many low-level events and logging them into a trace log can lead to issues, such as events being logged out of order or some events being incorrectly logged, and errors during script execution.

\textbf{Suggestions: }The key issue here is how to balance the resource of I/O operations with the comprehensiveness of semantic information acquisition. For the resource aspect, optimizing the efficiency and performance of the instrumenting tool and logging tool, and continuously monitoring resource usage on the testing device, can help alleviate the occurrence of such issues. For instance, optimizing the logging mechanism by using buffers or more efficient logging libraries to ensure that events are logged efficiently and correctly. For the information aspect, removing redundant information and retaining relevant semantic information can alleviate the problem of resource consumption. For example, by modifying the configuration to reduce the frequency of logging low-level events, the tool can alleviate the strain on I/O operations and reduce the likelihood of related issues. Prioritizing the logging of critical events or filtering out less important ones to reduce the overall volume of events being logged may also help.

\subsubsection{Screen understanding}
To generate replay scripts, vision-based \recordreplay tools need to calculate the coordination of actions by pixels. It is easy to raise issues that fail to generate scripts due to the required resolution or the first frame of the recording videos does not meet the default requirements. Furthermore, vision-based tools rely on the touch indicator to identify the type and location of the action. When the identified action is a long tap or a click, misjudgment is prone to occur. At the same time, as the Android touch indicator is white, it is difficult for V2S to identify the touch indicator in a white background, resulting in the generation of inaccurate scripts and click failed issues.

\textbf{Suggestions: }For resolution issues, adding dynamic mechanisms to adjust the coordination of generated actions based on the video resolution instead of using fixed settings can help solve this issue. To better locate the touch indicator and identify the action type, the following changes can be considered: (1) modifying the touch indicator color to increase contrast against the background; (2) introducing patterns or textures to the touch indicator, thereby making it distinct from the surrounding background and easy to recognize by the tool; (3) implementing a dynamic touch indicator that changes\Space{ color or appearance} when applying different actions, especially for long taps. 

\subsubsection{Flakiness in recording phase}
\sara involves the \emph{candidate recognizer} module, which can recognize the widgets under interactions and generate a unique identifier for each of the widgets so that they can be uniquely recognized in the replaying phase. However, this module cannot recognize widgets if the UI hierarchy is dynamically loaded, and will instead just wait for some time before choosing one of the candidate widgets. Therefore if the UI hierarchy is not statically loaded, then the chosen widget can be wrong, leading to click failed issues in the replaying phase.

\textbf{Suggestions: } A dynamic waiting and loading check mechanism to detect if the UI hierarchy is fully loaded can help.

\section{RQ$_3$: Bug reproduction from \aig tools}
\label{sec:rq3}

In this section, we explore the reproducibility of crashes detected by \aig tools. To reproduce such \aig detected bugs, developers can (1) rely on the seed functionality built-in to \aig tools, (2) improve \aig tools to reproduce crashes, and (3) use \aig tools with \recordreplay tools. We conduct experiments for each of these three methods and analyze their performance. 

\textbf{\aig Tools. }We consider all \aig tools from the Themis dataset and use three state-of-the-art tools: Monkey, \ape, and Humanoid.
\textit{Monkey}~\cite{AndroidMonkey} is an Android UI testing tool developed by Google and is one of the earliest \aig tools. 
The tool produces purely randomized UI event sequences and injects them into the target Android system without considering the design details of the app under test.
\emph{\ape} is a model-based approach that dynamically optimizes a model of the GUI of a given app by leveraging the runtime information during exploration.
\ape then uses the model to systematically generate events to explore the app.
\emph{Humanoid} uses deep learning (DL)-based techniques to learn experiences from app usage traces generated by humans. It will continually use a model-based policy to discover the app with several potential inputs according to the current state.

We exclude the other tools from the Themis dataset for various reasons.
Specifically, we exclude ComboDroid~\cite{combodroid} because when we use it in conjunction with \recordreplay tools, it is unable to run for long\Space{longer than \Num{10} minutes} due to incompatibilities.
For example, when we combine \sara with ComboDroid, we find that ComboDroid will frequently cause the process ID of an app to change.
However, \sara relies on attaching itself to the process ID of an app to record and replay actions, therefore \sara will stop recording a few minutes into a given testing session.
We omit TimeMachine~\cite{dong2020time} because it requires app instrumentation, which can conflict with \recordreplay tools.

\subsection{Built-in functionality of \aig tools}
To reproduce bugs detected by \aig tools themselves, we try to remove the randomness in the action sequence generation process from \aig tools by fixing the random seeds. As Humanoid is a DL-based technique, we cannot fully remove the randomness in exploration. We run the \aig tools on each app two times and compare the detected crashes. 
To better compare the results of the two runs, we use the following four metrics in our study:

\noindent \textbf{Overlap}: Are the exceptions in the runs overlapping?

\noindent \textbf{Unique}: Are the exception sets consistent across the runs?

\noindent \textbf{Order}: Is the order in which each exception is detected consistent between the two runs?

\noindent \textbf{Time}: Is the timing of each exception consistent in the runs?

The results are shown in Table~\ref{tab:rerun_table}.
Even if we use the same action sequences to run \aig tools repeatedly, it is still impossible to get completely consistent traces. If we relax the requirement to reproduce the same list of unique bugs in multiple runs, the success rate increases. Monkey, \ape and Humanoid can reproduce all detected bugs in seven, six, and two of the 27 apps, respectively.

\begin{table}[t]
  \centering
  \caption{Comparison of exception consistency across two runs of AIG tools on 27 apps}
\begin{tabular}{|l|l|r|r|r|r|}
\hline
\textbf{Tools}            & \textbf{Runs} & \multicolumn{1}{c|}{\textbf{Overlap}} & \multicolumn{1}{c|}{\textbf{Unique}} & \multicolumn{1}{c|}{\textbf{Order}} & \multicolumn{1}{c|}{\textbf{Time}} \\ \hline
\multirow{3}{*}{APE}      & Same              & 21                                    & 6                                    & 2                                   & 0                                  \\
                          & Not Same          & 3                                     & 18                                   & 22                                  & 24                                 \\
                          & NA                & 3                                     & 3                                    & 3                                   & 3                                  \\ \hline
\multirow{3}{*}{Monkey}   & Same              & 18                                    & 7                                    & 2                                   & 2                                  \\
                          & Not Same          & 3                                     & 14                                   & 19                                  & 19                                 \\
                          & NA                & 6                                     & 6                                    & 6                                   & 6                                  \\ \hline
\multirow{3}{*}{Humanoid} & Same              & 9                                     & 2                                    & 0                                   & 0                                  \\
                          & Not Same          & 7                                     & 14                                   & 16                                  & 16                                 \\
                          & NA                & 11                                    & 11                                   & 11                                  & 11                                 \\ \hline
\end{tabular}
\label{tab:rerun_table}
\vspace{-1em}
\end{table}

\subsection{Improved version of \aig tools}
In this experiment, we perform simple changes to \aig tools to understand whether such changes can help \aig tools reproduce crashes. Considering that \aig tools normally apply short action intervals (e.g., Monkey and \ape use 200ms by default), we conducted comparative experiments to measure the impact of different action intervals. We used action intervals of 500, 1000, and 2000 ms to repeat the experiments of Table~\ref{tab:rerun_table} with the same random seed. The results indicate that even with increased action intervals, there is no substantial improvement in the outcomes (e.g., up to 70\% of cases could not be reproduced for Monkey when the same seed is used). The detailed experiment results is in our artifact~\cite{appendix}.

\subsection{Combination of \aig and \recordreplay tools}
We explore how the state-of-the-art \recordreplay tools work in conjunction with \aig tools to reproduce \aig detected bugs. 

\subsubsection{Compatibility problems between \recordreplay and \aig tools} We find that \reran, \replaykit, and V2S are not compatible with \aig tools due to the following reasons.

\textbf{Android SDK conflicts.} \reran and \replaykit rely on the Android SDK \textit{getevent} tool to monitor and record all registered UI actions, while all \aig tools utilize the Android SDK \textit{sendevent} tool to operate actions. Due to limitations within the Android SDK mechanism, \textit{sendevent} and \textit{getevent} tools cannot function simultaneously. As a result, the studied \recordreplay tools cannot record the actions applied by \aig tools.

\textbf{Missing touch indicator.} V2S requires the recording videos to have the touch indicator of each performed action displayed on the screen, but \aig tools do not trigger touch indicator animations\Space{ when applying actions}. Moreover, generating scripts from videos in V2S incurs a high time cost, which is impractical as most \aig tools are expected to run for a  long time.

\begin{figure}[t]
    \centering
    \begin{subfigure}[b]{0.28\columnwidth}
        \includegraphics[width=\textwidth]{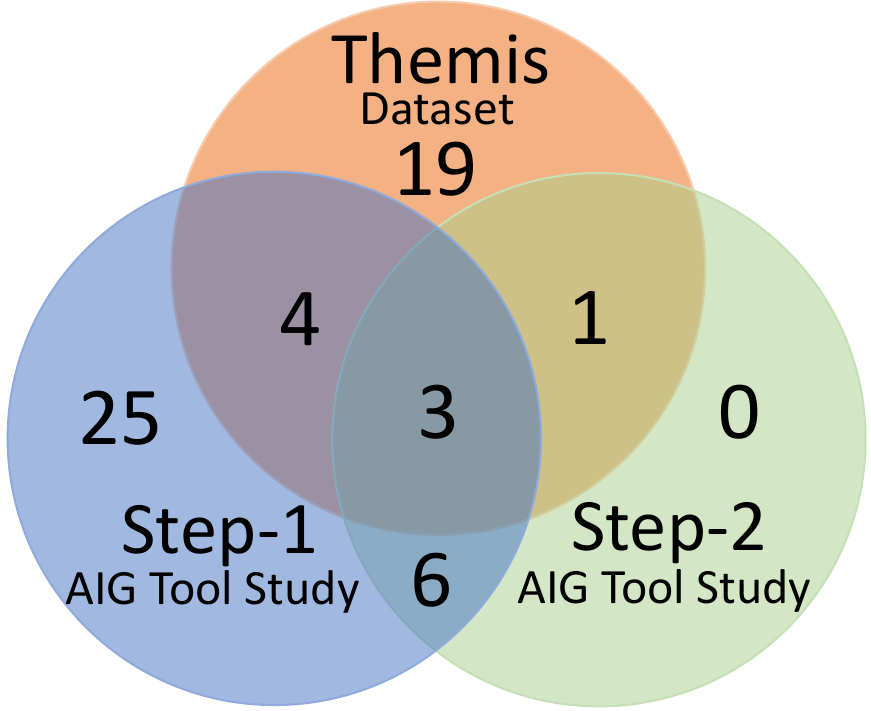}
        \caption{Monkey}
        \label{fig:monkey}
    \end{subfigure}
    \hfill
    \begin{subfigure}[b]{0.28\columnwidth}
        \includegraphics[width=\textwidth]{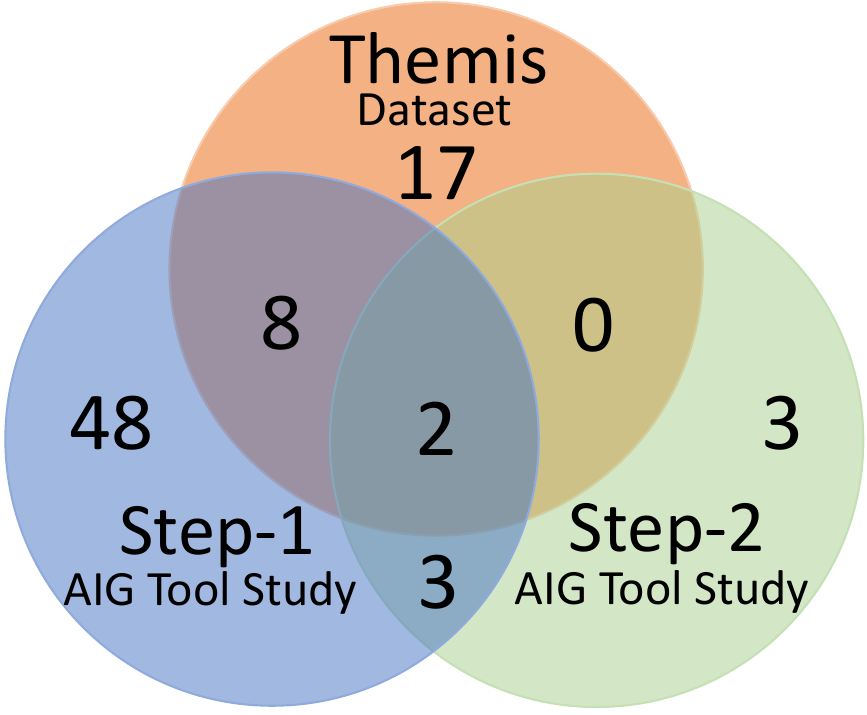}
        \caption{\ape}
        \label{fig:ape}
    \end{subfigure}
    \hfill
    \begin{subfigure}[b]{0.28\columnwidth}
        \includegraphics[width=\textwidth]{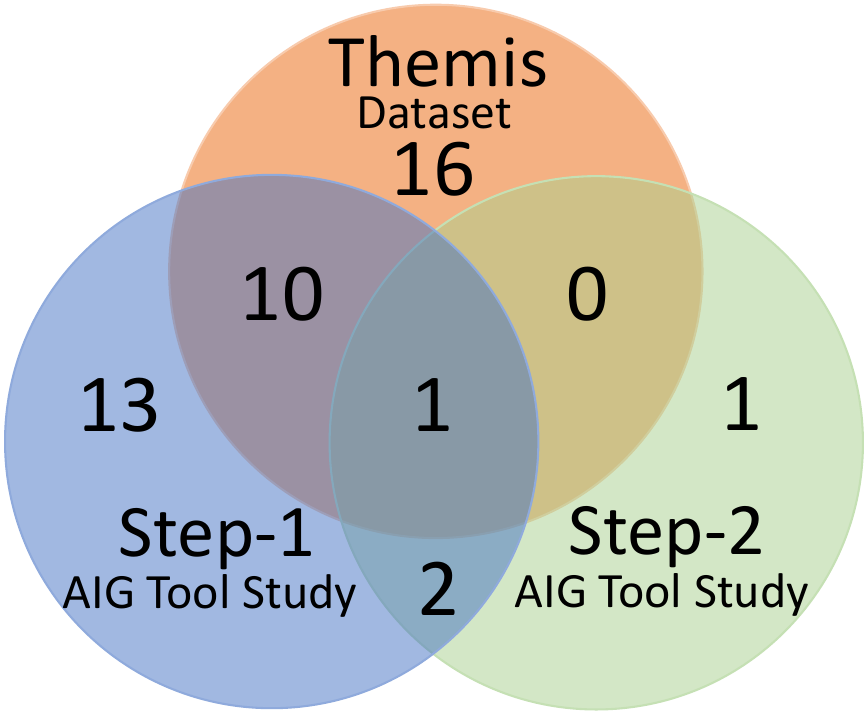}
        \caption{Humanoid}
        \label{fig:humanoid}
    \end{subfigure}
    \vspace{-1ex}
    \caption{Number of unique crashes detected by the Themis dataset and Steps 1 and 2 of our \aig Tool experiments.}
    \label{fig:aig:venn}

\end{figure}

\begin{table}[t]
  \caption{Breakdown of the types of crashes recorded when we combine \sara with \aig tools.}
  \vspace{-1ex}
  \centering
\begin{tabular}{|l|r|r|r|}
\hline
\textbf{Crash Type} & \textbf{\ape} & \textbf{Monkey} & \textbf{Humanoid}\\ \hline
Themis crash        & 2            & 4         & 1     \\
New crash           & 3            & 6        & 2     \\ \hline
\textbf{Total}               & \textbf{5}           & \textbf{10} & \textbf{3}              \\ \hline
\end{tabular}
\label{tab:aig_stage2}
\end{table}
\begin{table}[t]
  \centering
  \caption{Breakdown of the crashes we are unable to replay when we combine \sara with \aig tools.}
\begin{tabular}{|ll|r|r|r|}
\hline
\multicolumn{1}{|l|}{\textbf{Results}}                 & \textbf{Reason}     & \multicolumn{1}{c|}{\textbf{Ape}} & \multicolumn{1}{c|}{\textbf{Monkey}} & \multicolumn{1}{c|}{\textbf{Humanoid}} \\ \hline
\multicolumn{1}{|l|}{\multirow{4}{*}{\textbf{Fail}}} & Inaccurate action    & 2                                 & 4                                    & -                                      \\
\multicolumn{1}{|l|}{}                                 & Parsing Log Failed  & -                                 & 3                                    & 3                                      \\
\multicolumn{1}{|l|}{}                                 & Script Error        & -                                 & 2                                    & -                                      \\ \cline{2-5} 
\multicolumn{1}{|l|}{}                                 & \textbf{Total}      & \textbf{2}                        & \textbf{9}                           & \textbf{3}                             \\ \hline
\multicolumn{2}{|l|}{\textbf{Succeed}}                                       & 2                                 & 1                                    & 0                                      \\ \hline
\multicolumn{2}{|l|}{\textbf{Flaky}}                                         & 1                                 & 0                                    & 0                                      \\ \hline
\multicolumn{2}{|l|}{\textbf{Total}}                                         & \textbf{5}                        & \textbf{10}                          & \textbf{3}                             \\ \hline
\end{tabular}
\label{tab:aig_stage3}
\end{table}

\subsubsection{Unsuccessful cases in the combination of \sara and \aig tools} 
Our experiments find that Monkey, \ape and Humanoid detects \numMonkeyDetected, \numApeDetected, and \numHumanoidDetected crashes, respectively in Step~1 (described in Section~\ref{sec:methodology}).
To evaluate \sara{}'s recordability of the crashes that Monkey, \ape, and Humanoid detects, we combine the \aig tools with \sara in Step~2.
Fig.~\ref{fig:aig:venn} shows the breakdown of how the crashes we find from Step~2 compare to the crashes we find from Step~1 and the crashes in the Themis dataset. 
Overall, the majority of the recorded crashes from Monkey, \ape, and Humanoid are detected in Step 1 or from the Themis dataset.
To ensure that the crashes found in Step~2 are not introduced by \sara, we only keep the intersection of Step~2 with the Themis crashes and new crashes detected in Step 1. 
As shown in Table~\ref{tab:aig_stage2}, we successfully recorded \numMonkeyRecorded, \numApeRecorded, and \numHumanoidRecorded crashes detected by Monkey, \ape, and Humanoid, respectively in Step~2. Then we replayed these recorded crashes five times each. The replaying results are shown in Table~\ref{tab:aig_stage3} and described in Section~\ref{sec:limitation:replay}.
Overall, the current performance of record and replay tools in reproducing the crashes detected by \aig tools is far from satisfactory.

\textbf{Suggestion. }SDK conflict is the main problem of combining \aig and \recordreplay tools. To help, we suggest (1)~developing \recordreplay tools that can record events through other mechanisms and (2)~modifying Android SDK logic to remove existing conflicts\Space{ can improve this problem}.

\section{RQ$_4$: Root causes of issues in combining \aig and \recordreplay tools}


\subsection{Limitations in recording phase}
As mentioned before, when using \sara to record crashes detected by \aig tools, the number of recorded crashes is limited. The main reason is that \sara has the limitation of relying on process IDs~(PIDs) to record actions. During both the recording and replaying phases, PIDs must be provided as input to the program. However, when a crash is triggered, the app will restart and will be assigned a new PID. Once the PID is changed, \sara will lose the connection to the app and be unable to record any new actions. Therefore, in our experiment, we are able to record only the first crash for each run, because (1)~\sara must start later than the app and earlier than \aig tools to get the app PID as input to not miss actions, and (2)~we cannot interrupt the \aig tool's exploration in the middle and restart it as \aig tools need to maintain their exploration history to guide future exploration. A restart will lose the exploration history and potentially cause the tools to waste time exploring UI space that has already been explored.

\subsection{Limitations in replaying phase} 
\label{sec:limitation:replay}
As shown in Table~\ref{tab:aig_stage3}, \sara cannot replay most of the recorded crashes detected by the three \aig tools. The root causes can be categorized into the following three types.

\subsubsection{Inaccurate action} The actions performed by Monkey are not correctly recorded by \sara, therefore \sara ended up triggering a different crash than the expected crash. Listing~\ref{fig:actioncode} shows an example.
During the recording of the crash, \sara incorrectly records an extra swipe, which eventually triggers a different crash than the crash that was recorded. When we analyze the log files, we find that \sara will record all actions between ACTION\_DOWN and ACTION\_UP and consider them as one action, then split tap and swipe events by checking if the sub-sequence contains ACTION\_MOVE. During the recording process, this action appears as an unsuccessful swipe, it swipes slightly from the bottom right to the top left but fails to change the current page. In the trace recorded by \sara, this action is decomposed into three events, i.e., an extra \texttt{ACTION\_MOVE} is inserted between \texttt{ACTION\_DOWN} and \texttt{ACTION\_UP}. We believe that the root cause of this type of issue is consistent with the reasons described in Section~\ref{sec:sara-io}, which is due to logging issues caused by I/O process overload.

\textbf{Suggestion. }As mentioned in Section~\ref{sec:sara-io}, balancing the resource of I/O operations and the comprehensiveness of semantic information acquisition can help improve this issue.



\begin{figure}[t]
  \centering
  \begin{lstlisting}[style=shell, caption={Divergence action between recorded action sequence and the replayed action sequence.}, label={fig:actioncode}]
# Original events generated by Monkey
ACTION_DOWN: (796.0,826.0)
ACTION_UP: (703.7517,694.93243)

# Events recorded by SARA
ACTION_DOWN: (796.0,826.0)
ACTION_MOVE: (712.4744262695312, 714.1114501953125)
ACTION_UP: (703.7517,694.93243)
  \end{lstlisting}
  \vspace{-2em}
\end{figure}

\subsubsection{Parsing log failure} Monkey and Humanoid each have three cases that fall into this category. The error can come from two steps. Some cases happen before the normal replaying phase due to anomalous log writing. The output of the \sara recording process is the trace logs including all recorded actions. Before replaying the traces, \sara needs to transform the trace log into automated replay scripts. This error occurs when the transformation fails. 
When \sara tries to obtain the app activity hierarchy information, it will access the related XML files, which store the activity hierarchy data. If these XML files are empty, then \sara will raise a TypeError exception.
We also find one case that has flaky results when we replay the crash detected by \ape with \sara.  Specifically, WordPress\#10302~\cite{wordpress10302} cannot pass in all five runs due to two different reasons: (1)~\sara fails to replay the crash due to the parsing log failing three times, (2)~in the other two runs, the replay is unsuccessful due to missing actions.

\textbf{Suggestion. } (1)~Introducing log validation steps during the recording process to monitor in real-time whether there are any anomalies in the logs and correcting such anomalies or (2)~implementing more robust logging mechanisms to ensure that all necessary information is captured accurately during recording can help.

\subsubsection{Script error} In rare cases, the replay scripts generated by \sara raise errors. One use case is due to incomplete action records caused by the Frida tool blocking during the recording process. The other unsuccessful case is that \sara failed to get the activity information from Android's activity hierarchy tree. The tested app Openlauncher~\cite{openlauncher67} is a launcher app for Android, which is located on a different layer of the activity hierarchy tree compared with general apps. When \sara tries to parse the launcher app from the activity hierarchy tree, the different layer prevents \sara from getting the app information, and an indentation exception is raised. However, after fixing this indentation issue from \sara, the replaying is still unsuccessful due to missing actions in the record trace.

\textbf{Suggestion. }(1)~Involving dynamic layer detection functionality to automatically adjust the parsing logic based on the layer the app is located on or (2)~as mentioned in Section~\ref{sec:sara-io}, enlarging the action interval and improving the logging mechanisms can help alleviate missing action issues.

\section{Threats to Validity}
\label{sec:threats}

\noindent\textbf{Construct Validity. }Threats to construct validity concern the operationalization of experimental artifacts. In this study, we make use of past tools, datasets, and benchmarks. To mitigate potential threats to construct validity, we followed the instructions to instantiate and use both the \recordreplay tools and the \aig tools as described in the original papers and their corresponding online repositories/appendices. As mentioned earlier in the paper, we ran into issues in integrating \recordreplay and \aig tools due to technical limitations regarding how events are recorded. It is possible that such issues could be mitigated through additional engineering and integration efforts, however, this observed difficulty in tool use would also be faced by developers aiming to combine \aig and \recordreplay tools in a scenario similar to the one in our study.

\noindent\textbf{Internal Validity.}Threats to internal validity concern unexpected factors in the experiments that may contribute to observed findings. One of the main threats to the internal validity of our study stems from the filtering of our studied bugs, as the dataset utilized could ultimately lead to different results if filtered in different ways. We used two well-known artifacts, AndroR2 and the Themis benchmark, as our source of bugs. 
For the manual study, we employed four authors to reproduce each bug to confirm that the bugs were indeed reproducible and excluded non-reproducible bugs from our analysis (Step 1). Before reproducing each use case, all four authors discuss and agree on guidelines to be followed in recording each of the bugs (e.g., the steps in the corresponding videos for each Themis bug or the ``steps to reproduce'' and scripts in each AndroR2 bug report).
Each tool was then used by a specific author to record and replay reproducible bugs.
Differences in how authors may have attempted to record bugs could influence our results.
We mitigate this threat by discussing among authors what methodology should be used to record and replay each bug. All unsuccessful cases for each tool are then cross-checked by the other authors to ensure that all unsuccessful cases were not due to user bias.

For the \aig tool study, we used only those bugs that were detectable by Monkey, \ape, or Humanoid and filtered bugs we could not detect without \recordreplay tools. Another potential threat to internal validity is the analysis performed to root-cause the reasons for the lack of recordability or replayability. However, we again had multiple authors inspect these cases of interest to derive a mutually agreed-upon set of root causes. As a result of this methodology, we believe that we have sufficiently mitigated any threats to internal validity.

\noindent\textbf{External Validity.} Threats to external validity concern the generalizability of our results. While we cannot claim that our results generalize outside of our studied tools, apps, and bugs, we took several steps to ensure these experimental constructs were as generalizable as possible. AndroR2 is currently the largest dataset of non-crashing Android bugs in the research literature. Themis is a recent work that represents the largest set of reproducible crashing bugs and experimental infrastructure to run \aig tools. Given the diversity among the apps and bugs contained within these datasets, we believe that the results of this study are generalizable to the point where reasonably broad conclusions can be drawn. Another threat to external validity is the set of \recordreplay and \aig tools used. However, our selected tools represent state-of-the-art techniques across both academia and industry and provide a reasonably representative set of the current state of \aig and \recordreplay tools.

\section{Related work}
\label{sec:related}

\subsection{Record \& Replay Testing Tools for Android}

In the early stage of Android development, record and replay tools~\cite{reran, mobiplay, VALERA} made reasonable assumptions based on the characteristics of the era (e.g., simpler apps, fewer app updates, and fewer Android device models).
At this stage, tools are built based on coordinate representation without any search strategy. 
RERAN~\cite{reran} was a representative work at that time. When recording, RERAN captures user-input events and the coordinate information of the event by reading the screen device driver file at the kernel layer. Given the simplicity of the apps at that time, RERAN assumes that the apps are the same during playback as it was recorded, so the recorded coordinate information is directly written to the screen driver file to complete the replay.

As Android apps became more complex, the app updates became more frequent and device models became more fragmented, the techniques based on coordinate representation could handle cross-device record and replay. 
For this reason, techniques (such as SARA ~\cite{sara} and RANDR ~\cite{RANDR}) based on UI widgets and hierarchy attributes (with a certain extent of abstraction) were proposed.
For recording, SARA uses self-replay to record the motion events based on widgets, while RANDR uses only instrumentation. 
For replaying, both SARA and RANDR assume that the UI will not be changed, so they will just simply aim to perform the same events on the UI.
Much empirical evidence suggests such techniques are effective for cross-device record and replay given that the versions of the apps during recording and replaying are exactly the same~\cite{sara, RANDR, vet, glib, UIFlaky, wang2018empirical, zheng2017automated, malek19}. 

It is worth noting that although coordinate-based techniques do not work for cross-device record and replay, RERAN is still the most effective approach for same-device record and replay~\cite{lam17:record}. 
That being said, RERAN can be intrusive to the system (needs to be operated at the kernel layer to use the screen driver), but V2S~\cite{V2S} and RoScript~\cite{roscript} need to record and parse video to get the coordinates, still limits the practicality of these two tools.
\Space{but the performance of both is limited by the capabilities of current deep learning techniques.}


\subsection{Automated Input Generation (AIG) Tools for Android}

\subsubsection{Random-based Techniques}
Random testing is a software testing technique where programs are tested by generating random, independent inputs.
The most representative random-based technique is the Monkey~\cite{AndroidMonkey} tool included in Android SDK.
Monkey can send pseudo-random user events (e.g., key presses, screen taps) to the device.
Traditional work such as Monkey either focuses on event-based input such as UI events and system events, or relies on manual effort to generate valid text input. Dynodroid~\cite{machiry2013dynodroid} and Mulliner et al.~\cite{mulliner2009fuzzing} expand the UI events with system events, such as SMS receiving. However, when encountering a text input field like password, they must pause the testing and wait for manual input. 
More recent work generates testing input based on domain knowledge about the input structures. For example, prior works~\cite{ye2013droidfuzzer, sasnauskas2014intent,wu2016crafting,feng2016understanding} focus on fuzzing critical data structures in Android such as Intent and Binder, which are well-documented.  
In addition, another thread of work like Caiipa~\cite{liang2014caiipa} uses synthesized context observed in the wild to guide its fuzzer so that it can cover different context variations. 

\subsubsection{Model-based Techniques}
Model-based testing is one of the most common methodologies used in Android testing~\cite{azim2013targeted, li2017droidbot, arnatovich2016achieving, rastogi2013appsplayground, zhang2023scene, chen2019storydroid, hu2024enhancing}.
Before any model-based testing tool was developed specifically for testing Android apps, Takala et al.~\cite{takala2011experiences} performed experiments using TeMA~\cite{TEMA} which was originally developed for testing Series 60 (running on top of Symbian OS) GUI apps.
ORBIT~\cite{ORBITYang} is the first model-based Android testing technique based on a combination of dynamic GUI crawling and static code analysis, using analysis to avoid the generation of irrelevant UI events.

\subsubsection{Learning-based Techniques}
Some modern works~\cite{mao2016sapienz, vuong2018reinforcement, li2019humanoid, pan2020reinforcement, collins2021deep} generate inputs by leveraging machine learning techniques (such as neural nets) or search-based techniques (such as genetic algorithms). 

Another thread of work by Liu et al.~\cite{liu2017automatic} utilizes RNN to train a learning model and use it to generate text input based on the app context. Unfortunately, it requires a large amount of manual effort to write the training input.

\subsubsection{Widget-based Techniques}
UI widget identification is often combined and used together with app testing because an exerciser needs to interact with different UI widgets.  Super~\cite{SUPOR} and UIPicker~\cite{UIPicker} extract UI widget information from the layout's XML file and then identify sensitive inputs. UiRef~\cite{UiRef} improves prior works by adopting a hybrid approach that combines both static and dynamic identifications: the static method identifies widgets from layouts, just like prior work and the dynamic method extracts each rendered layout during on-device execution. Similarly,  CuriousDroid~\cite{CuriousDroid} instruments the Dalvik virtual machine to obtain the UI widgets and generate UI-related events during execution.

\subsubsection{Flaky Tests}

One common use case for \recordreplayFull tools is to record a user scenario of an app and every time the app changes, these common user scenarios are replayed to ensure that recent changes did not break existing functionality.
This use case of \recordreplay tools is similar to regression testing\Space{ for non-mobile software development}.
In the topic of regression testing, one common problem is known as flaky tests~\cite{Luo2014HEM, Eck2019PCB, Labuschagne2017IH}, which are tests that can non-deterministically pass or fail even on the same version of code. 
Flaky tests can arise if tests depend on specific thread interleaving~(concurrency)~\cite{Dong2021TYR}, test execution orders~\cite{Zhang2014JWMLEN, Lam2019OSMX, Bell2014K, Gyori2015SHM, Gruber2021LKF}, unspecified implementation details~\cite{Shi2016GLM}, time of the day, network call result, etc.
Similar to flaky tests, \recordreplay tools may sometimes succeed in replaying a user scenario and fail some other times even when the version of the app remains the same.
In this paper, we present a detailed analysis of the flakiness in replay scenarios for both manually recorded traces from bug reports and automatically recorded traces from \aig tools.

\section{Conclusion}
\label{sec:conclusion}

In this paper, we have presented the first comprehensive study examining the recordability and replayability of Android bug reproduction by four state-of-the-art \recordreplay tools. In particular, we conducted two studies aimed at recording and replaying manually reproduced common user scenarios, non-crashing bugs and crashes, and automatically discovered crashing bugs (via \aig tools). Our results illustrate that \perfailure{} of non-crashing bugs, \percrash{} of crashing bugs, and \perscenario{} of common user scenarios were not able to be reliably replayed, with the most prevalent underlying reasons for non-replayability being small action intervals and tool limitations in recording and inability to capture certain types of UI interactions. Moreover, existing methods are still far away from satisfactory to reliably reproduce crashes detected by \aig tools. Our findings highlight important current limitations with both \recordreplay and \aig tools that serve as impediments to the practical adoption of our studied tools and important directions for future research.

\section*{Acknowledgments}
We acknowledge
NSF grants CCF-2146443, CCF-2338287\OurComment{Wing's CAREER}, CNS-2235137, and the Amazon Trust AI Research Award, as well as Dragon Testing Technology for their support.

\bibliographystyle{IEEEtran}
\bibliography{ref}

\begin{thebibliography}{10}
\providecommand{\url}[1]{#1}
\csname url@samestyle\endcsname
\providecommand{\newblock}{\relax}
\providecommand{\bibinfo}[2]{#2}
\providecommand{\BIBentrySTDinterwordspacing}{\spaceskip=0pt\relax}
\providecommand{\BIBentryALTinterwordstretchfactor}{4}
\providecommand{\BIBentryALTinterwordspacing}{\spaceskip=\fontdimen2\font plus
\BIBentryALTinterwordstretchfactor\fontdimen3\font minus \fontdimen4\font\relax}
\providecommand{\BIBforeignlanguage}[2]{{%
\expandafter\ifx\csname l@#1\endcsname\relax
\typeout{** WARNING: IEEEtran.bst: No hyphenation pattern has been}%
\typeout{** loaded for the language `#1'. Using the pattern for}%
\typeout{** the default language instead.}%
\else
\language=\csname l@#1\endcsname
\fi
#2}}
\providecommand{\BIBdecl}{\relax}
\BIBdecl

\bibitem{reran}
L.~Gomez, I.~Neamtiu, T.~Azim, and T.~Millstein, ``{RERAN}: Timing-and touch-sensitive record and replay for {Android},'' in \emph{Proceedings of the 35th {IEEE}/{ACM} International Conference on Software Engineering ({ICSE})}, 2013, pp. 72--81.

\bibitem{VALERA}
Y.~Hu, T.~Azim, and I.~Neamtiu, ``Versatile yet lightweight record-and-replay for {Android},'' in \emph{Proceedings of the 2015 {ACM} {SIGPLAN} International Conference on Object-Oriented Programming, Systems, Languages, and Applications ({OOPSLA})}, 2015, pp. 349--366.

\bibitem{V2S}
M.~Havranek, C.~Bernal-C{\'a}rdenas, N.~Cooper, O.~Chaparro, D.~Poshyvanyk, and K.~Moran, ``{V2S}: A tool for translating video recordings of mobile app usages into replayable scenarios,'' in \emph{Proceedings of the 43rd {IEEE}/{ACM} International Conference on Software Engineering: Companion Proceedings ({ICSE}-Companion)}, 2021, pp. 65--68.

\bibitem{lam17:record}
W.~Lam, Z.~Wu, D.~Li, W.~Wang, H.~Zheng, H.~Luo, P.~Yan, Y.~Deng, and T.~Xie, ``Record and replay for {Android}: Are we there yet in industrial cases?'' in \emph{Proceedings of the 11th ACM Joint Meeting on Foundations of Software Engineering ({ESEC}/{FSE})}, 2017, pp. 854--859.

\bibitem{Su2021WS}
T.~Su, J.~Wang, and Z.~Su, ``Benchmarking automated {GUI} testing for {Android} against real-world bugs,'' in \emph{Proceedings of the 29th ACM Joint Meeting on European Software Engineering Conference and Symposium on the Foundations of Software Engineering ({ESEC}/{FSE})}, 2021, pp. 119--130.

\bibitem{liu2023understanding}
H.~Liu, Q.~Kong, J.~Wang, T.~Su, and H.~Sun, ``Understanding the reproducibility issues of {Monkey} for {GUI} testing,'' in \emph{International Symposium on Dependable Software Engineering: Theories, Tools, and Applications ({SETTA})}, 2023, pp. 132--151.

\bibitem{xiong2023empirical}
Y.~Xiong, M.~Xu, T.~Su, J.~Sun, J.~Wang, H.~Wen, G.~Pu, J.~He, and Z.~Su, ``An empirical study of functional bugs in {Android} apps,'' in \emph{Proceedings of the 32nd ACM SIGSOFT International Symposium on Software Testing and Analysis ({ISSTA})}, 2023, pp. 1319--1331.

\bibitem{sara}
J.~Guo, S.~Li, J.-G. Lou, Z.~Yang, and T.~Liu, ``{SARA}: Self-replay augmented record and replay for {Android} in industrial cases,'' in \emph{Proceedings of the 28th {ACM} {SIGSOFT} International Symposium on Software Testing and Analysis ({ISSTA})}, 2019, pp. 90--100.

\bibitem{Linares-Vasquez:ICSME'17}
M.~Linares-V{\'a}squez, C.~Bernal-C{\'a}rdenas, K.~Moran, and D.~Poshyvanyk, ``How do developers test {Android} applications?'' in \emph{Proceedings of the 2017 IEEE International Conference on Software Maintenance and Evolution ({ICSME})}.\hskip 1em plus 0.5em minus 0.4em\relax IEEE, 2017, pp. 613--622.

\bibitem{AndroidMonkey}
\BIBentryALTinterwordspacing
{Google}, ``{Android Monkey},'' 2022. [Online]. Available: \url{https://developer.android.com/studio/test/monkey}
\BIBentrySTDinterwordspacing

\bibitem{appendix}
\BIBentryALTinterwordspacing
Z.~Song, S.~M.~H. Mansur, R.~S. Rathnasuriya, Y.~Fatima, W.~Yang, K.~Moran, and W.~Lam, ``{Online Appendix},'' 2024. [Online]. Available: \url{https://recordreplaystudy.github.io/}
\BIBentrySTDinterwordspacing

\bibitem{appium}
\BIBentryALTinterwordspacing
{Appium}, ``{Appium}: Cross-platform test automation for native, hybrid, mobile web, and desktop apps,'' 2019. [Online]. Available: \url{https://github.com/appium/appium}
\BIBentrySTDinterwordspacing

\bibitem{culebra}
\BIBentryALTinterwordspacing
{Milano, Diego Torres}, ``{Culebra}: {GUI} tool for generating {AndroidViewClient} scripts,'' 2018. [Online]. Available: \url{https://github.com/dtmilano/AndroidViewClient/wiki/culebra}
\BIBentrySTDinterwordspacing

\bibitem{gifdroid}
S.~Feng and C.~Chen, ``{GIFDroid}: Automated replay of visual bug reports for {Android} apps,'' in \emph{Proceedings of the 44th {IEEE}/{ACM} International Conference on Software Engineering ({ICSE})}, 2022, pp. 1045--1057.

\bibitem{monkeyrunner}
\BIBentryALTinterwordspacing
{Google}, ``{monkeyrunner},'' 2022. [Online]. Available: \url{https://developer.android.com/studio/test/monkeyrunner}
\BIBentrySTDinterwordspacing

\bibitem{mosaic}
M.~Halpern, Y.~Zhu, R.~Peri, and V.~J. Reddi, ``{Mosaic}: Cross-platform user-interaction record and replay for the fragmented {Android} ecosystem,'' in \emph{Proceedings of the 2015 {IEEE} International Symposium on Performance Analysis of Systems and Software ({ISPASS})}, 2015, pp. 215--224.

\bibitem{Robotium}
\BIBentryALTinterwordspacing
{RobotiumTech}, ``{Robotium}: User scenario testing for {Android},'' 2021. [Online]. Available: \url{https://github.com/robotiumtech/robotium}
\BIBentrySTDinterwordspacing

\bibitem{replaykit}
\BIBentryALTinterwordspacing
{AppetizerIO}, ``{ReplayKit}: Command line tools for recording, replaying, and mirroring touchscreen events for {Android},'' 2019. [Online]. Available: \url{https://github.com/appetizerio/replaykit}
\BIBentrySTDinterwordspacing

\bibitem{frost}
\BIBentryALTinterwordspacing
A.~Wang, ``{Frost for Facebook}: An extensive and functional third-party app for {Facebook},'' 2022. [Online]. Available: \url{https://github.com/AllanWang/Frost-for-Facebook}
\BIBentrySTDinterwordspacing

\bibitem{firefox}
\BIBentryALTinterwordspacing
{Mozilla}, ``{Firefox Lite}: Emerging market experiment,'' 2021. [Online]. Available: \url{https://github.com/mozilla-mobile/FirefoxLite}
\BIBentrySTDinterwordspacing

\bibitem{androR2dataset}
\BIBentryALTinterwordspacing
{Fazzini, Mattia} and {Johnson, Jack}, ``{AndroR2+}: An empirical investigation into the reproduction of bug reports for {Android} apps,'' 2022. [Online]. Available: \url{https://github.com/se-umn/2022_saner_bug_report_reproduction_study}
\BIBentrySTDinterwordspacing

\bibitem{frida}
\BIBentryALTinterwordspacing
{Frida}, ``{Frida}: Dynamic instrumentation toolkit,'' 2019. [Online]. Available: \url{https://github.com/frida/frida}
\BIBentrySTDinterwordspacing

\bibitem{combodroid}
J.~Wang, Y.~Jiang, C.~Xu, C.~Cao, X.~Ma, and J.~Lu, ``{Combodroid}: Generating high-quality test inputs for {Android} apps via use case combinations,'' in \emph{Proceedings of the {ACM}/{IEEE} 42nd International Conference on Software Engineering ({ICSE})}, 2020, pp. 469--480.

\bibitem{dong2020time}
Z.~Dong, M.~B{\"o}hme, L.~Cojocaru, and A.~Roychoudhury, ``Time-travel testing of {Android} apps,'' in \emph{Proceedings of the {ACM}/{IEEE} 42nd International Conference on Software Engineering ({ICSE})}, 2020, pp. 481--492.

\bibitem{wordpress10302}
\BIBentryALTinterwordspacing
{Ercoli, Danilo}, ``Fix {NPE} on login screen when tapping help icon,'' 2019. [Online]. Available: \url{https://github.com/wordpress-mobile/WordPress-Android/pull/10302}
\BIBentrySTDinterwordspacing

\bibitem{openlauncher67}
\BIBentryALTinterwordspacing
{TheLastProject}, ``Crash when opening sound panel while do not disturb is activated,'' 2017. [Online]. Available: \url{https://github.com/OpenLauncherTeam/openlauncher/issues/67}
\BIBentrySTDinterwordspacing

\bibitem{mobiplay}
Z.~Qin, Y.~Tang, E.~Novak, and Q.~Li, ``{MobiPlay}: A remote execution-based record-and-replay tool for mobile applications,'' in \emph{Proceedings of the 38th {IEEE}/{ACM} International Conference on Software Engineering ({ICSE})}, 2016, pp. 571--582.

\bibitem{RANDR}
O.~Sahin, A.~Aliyeva, H.~Mathavan, A.~Coskun, and M.~Egele, ``Towards practical record-and-replay for mobile applications,'' in \emph{Proceedings of the 56th Annual Design Automation Conference ({DAC})}, 2019, pp. 1--2.

\bibitem{vet}
W.~Wang, W.~Yang, T.~Xu, and T.~Xie, ``{VET}: Identifying and avoiding {UI} exploration tarpits,'' in \emph{Proceedings of the 29th {ACM} Joint Meeting on European Software Engineering Conference and Symposium on the Foundations of Software Engineering ({ESEC}/{FSE})}, 2021, pp. 83--94.

\bibitem{glib}
K.~Chen, Y.~Li, Y.~Chen, C.~Fan, Z.~Hu, and W.~Yang, ``{GLIB}: Towards an automated test oracle for graphically-rich applications,'' in \emph{Proceedings of the 29th {ACM} Joint Meeting on European Software Engineering Conference and Symposium on the Foundations of Software Engineering ({ESEC}/{FSE})}, 2021, pp. 1093--1104.

\bibitem{UIFlaky}
A.~Romano, Z.~Song, S.~Grandhi, W.~Yang, and W.~Wang, ``An empirical analysis of {UI}-based flaky tests,'' in \emph{Proceedings of the 43rd {IEEE}/{ACM} International Conference on Software Engineering ({ICSE})}, 2021, pp. 1585--1597.

\bibitem{wang2018empirical}
W.~Wang, D.~Li, W.~Yang, Y.~Cao, Z.~Zhang, Y.~Deng, and T.~Xie, ``An empirical study of {Android} test generation tools in industrial cases,'' in \emph{Proceedings of the 33rd {ACM}/{IEEE} International Conference on Automated Software Engineering ({ASE})}, 2018, pp. 738--748.

\bibitem{zheng2017automated}
H.~Zheng, D.~Li, B.~Liang, X.~Zeng, W.~Zheng, Y.~Deng, W.~Lam, W.~Yang, and T.~Xie, ``Automated test input generation for {Android}: Towards getting there in an industrial case,'' in \emph{Proceedings of the 39th {IEEE}/{ACM} International Conference on Software Engineering: Software Engineering in Practice Track ({ICSE}-{SEIP})}, 2017, pp. 253--262.

\bibitem{malek19}
J.-W. Lin, R.~Jabbarvand, and S.~Malek, ``Test transfer across mobile apps through semantic mapping,'' in \emph{Proceedings of the 34th {IEEE}/{ACM} International Conference on Automated Software Engineering ({ASE})}, 2019, pp. 42--53.

\bibitem{roscript}
J.~Qian, Z.~Shang, S.~Yan, Y.~Wang, and L.~Chen, ``{Roscript}: A visual script-driven truly non-intrusive robotic testing system for touch screen applications,'' in \emph{Proceedings of the 42nd {ACM}/{IEEE} International Conference on Software Engineering ({ICSE})}, 2020, pp. 297--308.

\bibitem{machiry2013dynodroid}
A.~Machiry, R.~Tahiliani, and M.~Naik, ``{Dynodroid}: An input generation system for {Android} apps,'' in \emph{Proceedings of the 9th {ACM}/{IEEE} Joint Meeting on Foundations of Software Engineering ({ESEC}/{FSE})}, 2013, pp. 224--234.

\bibitem{mulliner2009fuzzing}
C.~Mulliner and C.~Miller, ``Fuzzing the phone in your phone,'' \emph{Black Hat {USA}}, vol.~25, p.~31, 2009.

\bibitem{ye2013droidfuzzer}
H.~Ye, S.~Cheng, L.~Zhang, and F.~Jiang, ``{DroidFuzzer}: Fuzzing {Android} apps with the intent-filter tag,'' in \emph{Proceedings of the International Conference on Advances in Mobile Computing \& Multimedia ({MoMM})}, 2013, pp. 68--74.

\bibitem{sasnauskas2014intent}
R.~Sasnauskas and J.~Regehr, ``{Intent fuzzer}: Crafting intents of death,'' in \emph{Proceedings of the 2014 Joint International Workshop on Dynamic Analysis ({WODA}) and Software and System Performance Testing, Debugging, and Analytics ({PERTEA})}, 2014, pp. 1--5.

\bibitem{wu2016crafting}
T.~Wu and Y.~Yang, ``Crafting intents to detect {ICC} vulnerabilities of {Android} apps,'' in \emph{Proceedings of the 12th International Conference on Computational Intelligence and Security ({CIS})}, 2016, pp. 557--560.

\bibitem{feng2016understanding}
H.~Feng and K.~G. Shin, ``Understanding and defending the {Binder} attack surface in {Android},'' in \emph{Proceedings of the 32nd Annual Conference on Computer Security Applications ({ACSAC})}, 2016, pp. 398--409.

\bibitem{liang2014caiipa}
C.-J.~M. Liang, N.~D. Lane, N.~Brouwers, L.~Zhang, B.~F. Karlsson, H.~Liu, Y.~Liu, J.~Tang, X.~Shan, R.~Chandra \emph{et~al.}, ``{Caiipa}: Automated large-scale mobile app testing through contextual fuzzing,'' in \emph{Proceedings of the 20th Annual International Conference on Mobile Computing and Networking ({MobiCom})}, 2014, pp. 519--530.

\bibitem{azim2013targeted}
T.~Azim and I.~Neamtiu, ``Targeted and depth-first exploration for systematic testing of {Android} apps,'' in \emph{Proceedings of the 2013 {ACM} {SIGPLAN} International Conference on Object-Oriented Programming, Systems, Languages \& Applications ({OOPSLA})}, 2013, pp. 641--660.

\bibitem{li2017droidbot}
Y.~Li, Z.~Yang, Y.~Guo, and X.~Chen, ``{DroidBot}: A lightweight {UI}-guided test input generator for {Android},'' in \emph{Proceedings of the 39th {IEEE}/{ACM} International Conference on Software Engineering Companion ({ICSE}-{C})}, 2017, pp. 23--26.

\bibitem{arnatovich2016achieving}
Y.~L. Arnatovich, M.~N. Ngo, T.~H.~B. Kuan, and C.~Soh, ``Achieving high code coverage in {Android} {UI} testing via automated widget exercising,'' in \emph{Proceedings of the 23rd Asia-Pacific Software Engineering Conference ({APSEC})}, 2016, pp. 193--200.

\bibitem{rastogi2013appsplayground}
V.~Rastogi, Y.~Chen, and W.~Enck, ``{AppsPlayground}: Automatic security analysis of smartphone applications,'' in \emph{Proceedings of the 3rd {ACM} Conference on Data and Application Security and Privacy ({CODASPY})}, 2013, pp. 209--220.

\bibitem{zhang2023scene}
X.~Zhang, L.~Fan, S.~Chen, Y.~Su, and B.~Li, ``Scene-driven exploration and {GUI} modeling for {Android} apps,'' in \emph{Proceedings of the 38th {IEEE}/{ACM} International Conference on Automated Software Engineering ({ASE})}, 2023, pp. 1251--1262.

\bibitem{chen2019storydroid}
S.~Chen, L.~Fan, C.~Chen, T.~Su, W.~Li, Y.~Liu, and L.~Xu, ``{StoryDroid}: Automated generation of storyboard for {Android} apps,'' in \emph{Proceedings of the 41st {IEEE}/{ACM} International Conference on Software Engineering ({ICSE})}, 2019, pp. 596--607.

\bibitem{hu2024enhancing}
H.~Hu, H.~Wang, R.~Dong, X.~Chen, and C.~Chen, ``Enhancing {GUI} exploration coverage of {Android} apps with deep link-integrated {Monkey},'' \emph{{ACM} Transactions on Software Engineering and Methodology ({TOSEM})}, 2024.

\bibitem{takala2011experiences}
T.~Takala, M.~Katara, and J.~Harty, ``Experiences of system-level model-based {GUI} testing of an {Android} application,'' in \emph{Proceedings of the 4th {IEEE} International Conference on Software Testing, Verification and Validation ({ICST})}, 2011, pp. 377--386.

\bibitem{TEMA}
\BIBentryALTinterwordspacing
M.~Katara, ``{TEMA} adapter for {Android},'' 2010. [Online]. Available: \url{https://github.com/tema-tut/tema-android-adapter}
\BIBentrySTDinterwordspacing

\bibitem{ORBITYang}
W.~Yang, M.~R. Prasad, and T.~Xie, ``A grey-box approach for automated {GUI}-model generation of mobile applications,'' in \emph{Proceedings of the International Conference on Fundamental Approaches to Software Engineering ({FASE})}, 2013, pp. 250--265.

\bibitem{mao2016sapienz}
K.~Mao, M.~Harman, and Y.~Jia, ``{Sapienz}: Multi-objective automated testing for {Android} applications,'' in \emph{Proceedings of the 25th {ACM} {SIGSOFT} International Symposium on Software Testing and Analysis ({ISSTA})}, 2016, pp. 94--105.

\bibitem{vuong2018reinforcement}
T.~A.~T. Vuong and S.~Takada, ``A reinforcement learning-based approach to automated testing of {Android} applications,'' in \emph{Proceedings of the 9th {ACM} {SIGSOFT} International Workshop on Automating Test Case Design, Selection, and Evaluation ({A-TEST})}, 2018, pp. 31--37.

\bibitem{li2019humanoid}
Y.~Li, Z.~Yang, Y.~Guo, and X.~Chen, ``{Humanoid}: A deep learning-based approach to automated black-box {Android} app testing,'' in \emph{Proceedings of the 34th {IEEE}/{ACM} International Conference on Automated Software Engineering ({ASE})}, 2019, pp. 1070--1073.

\bibitem{pan2020reinforcement}
M.~Pan, A.~Huang, G.~Wang, T.~Zhang, and X.~Li, ``Reinforcement learning-based curiosity-driven testing of {Android} applications,'' in \emph{Proceedings of the 29th {ACM} {SIGSOFT} International Symposium on Software Testing and Analysis ({ISSTA})}, 2020, pp. 153--164.

\bibitem{collins2021deep}
E.~Collins, A.~Neto, A.~Vincenzi, and J.~Maldonado, ``Deep reinforcement learning-based {Android} application {GUI} testing,'' in \emph{Proceedings of the 35th Brazilian Symposium on Software Engineering ({SBES})}, 2021, pp. 186--194.

\bibitem{liu2017automatic}
P.~Liu, X.~Zhang, M.~Pistoia, Y.~Zheng, M.~Marques, and L.~Zeng, ``Automatic text input generation for mobile testing,'' in \emph{Proceedings of the 39th {IEEE}/{ACM} International Conference on Software Engineering ({ICSE})}, 2017, pp. 643--653.

\bibitem{SUPOR}
J.~Huang, Z.~Li, X.~Xiao, Z.~Wu, K.~Lu, X.~Zhang, and G.~Jiang, ``{SUPOR}: Precise and scalable sensitive user input detection for {Android} apps,'' in \emph{Proceedings of the 24th USENIX Conference on Security Symposium (SEC)}, 2015, pp. 977--992.

\bibitem{UIPicker}
Y.~Nan, M.~Yang, Z.~Yang, S.~Zhou, G.~Gu, and X.~Wang, ``{UIPicker}: {User-Input} privacy identification in mobile applications,'' in \emph{Proceedings of the 24th USENIX Conference on Security Symposium (SEC)}, 2015, pp. 993--1008.

\bibitem{UiRef}
B.~Andow, A.~Acharya, D.~Li, W.~Enck, K.~Singh, and T.~Xie, ``{UiRef}: Analysis of sensitive user inputs in {Android} applications,'' in \emph{Proceedings of the 10th {ACM} Conference on Security and Privacy in Wireless and Mobile Networks ({WiSec})}, 2017, pp. 23--34.

\bibitem{CuriousDroid}
P.~Carter, C.~Mulliner, M.~Lindorfer, W.~Robertson, and E.~Kirda, ``{CuriousDroid}: Automated user interface interaction for {Android} application analysis sandboxes,'' in \emph{Financial Cryptography and Data Security: 20th International Conference ({FC} 2016), Christ Church, Barbados, February 22--26, 2016, Revised Selected Papers}, 2017, pp. 231--249.

\bibitem{Luo2014HEM}
Q.~Luo, F.~Hariri, L.~Eloussi, and D.~Marinov, ``An empirical analysis of flaky tests,'' in \emph{Proceedings of the 22nd {ACM} {SIGSOFT} International Symposium on Foundations of Software Engineering ({FSE})}, 2014, pp. 643--653.

\bibitem{Eck2019PCB}
M.~Eck, F.~Palomba, M.~Castelluccio, and A.~Bacchelli, ``Understanding flaky tests: The developer’s perspective,'' in \emph{Proceedings of the 27th {ACM} Joint Meeting on European Software Engineering Conference and Symposium on the Foundations of Software Engineering ({ESEC}/{FSE})}, 2019, pp. 830--840.

\bibitem{Labuschagne2017IH}
A.~Labuschagne, L.~Inozemtseva, and R.~Holmes, ``Measuring the cost of regression testing in practice: A study of {Java} projects using continuous integration,'' in \emph{Proceedings of the 11th {ACM} Joint Meeting on Foundations of Software Engineering ({ESEC}/{FSE})}, 2017, pp. 821--830.

\bibitem{Dong2021TYR}
Z.~Dong, A.~Tiwari, X.~L. Yu, and A.~Roychoudhury, ``Flaky test detection in {Android} via event order exploration,'' in \emph{Proceedings of the 29th {ACM} Joint Meeting on European Software Engineering Conference and Symposium on the Foundations of Software Engineering ({ESEC}/{FSE})}, 2021, pp. 367--378.

\bibitem{Zhang2014JWMLEN}
S.~Zhang, D.~Jalali, J.~Wuttke, K.~Mu{\c{s}}lu, W.~Lam, M.~D. Ernst, and D.~Notkin, ``Empirically revisiting the test independence assumption,'' in \emph{Proceedings of the 2014 International Symposium on Software Testing and Analysis ({ISSTA})}, 2014, pp. 385--396.

\bibitem{Lam2019OSMX}
W.~Lam, R.~Oei, A.~Shi, D.~Marinov, and T.~Xie, ``{iDFlakies}: A framework for detecting and partially classifying flaky tests,'' in \emph{Proceedings of the 12th {IEEE} Conference on Software Testing, Validation and Verification ({ICST})}, 2019, pp. 312--322.

\bibitem{Bell2014K}
J.~Bell and G.~Kaiser, ``Unit test virtualization with {VMVM},'' in \emph{Proceedings of the 36th {IEEE}/{ACM} International Conference on Software Engineering ({ICSE})}, 2014, pp. 550--561.

\bibitem{Gyori2015SHM}
A.~Gyori, A.~Shi, F.~Hariri, and D.~Marinov, ``Reliable testing: Detecting state-polluting tests to prevent test dependency,'' in \emph{Proceedings of the 2015 ACM SIGSOFT International Symposium on Software Testing and Analysis ({ISSTA})}, 2015, pp. 223--233.

\bibitem{Gruber2021LKF}
M.~Gruber, S.~Lukasczyk, F.~Kroi{\ss}, and G.~Fraser, ``An empirical study of flaky tests in {Python},'' in \emph{Proceedings of the 14th {IEEE} Conference on Software Testing, Verification and Validation ({ICST})}, 2021, pp. 148--158.

\bibitem{Shi2016GLM}
A.~Shi, A.~Gyori, O.~Legunsen, and D.~Marinov, ``Detecting assumptions on deterministic implementations of non-deterministic specifications,'' in \emph{Proceedings of the 2016 {IEEE} International Conference on Software Testing, Verification and Validation ({ICST})}, 2016, pp. 80--90.

\end{thebibliography}

\end{document}